\documentclass[final,5p,times,twocolumn]{elsarticle}

\usepackage{amssymb,amsmath,amsfonts}

\usepackage{xcolor}

\usepackage{algorithmicx}
\usepackage{algpseudocode}
\usepackage{algorithm}

\usepackage{mathtools}

\usepackage{xpatch}
\makeatletter
\xpatchcmd\ALG@step{\arabic{ALG@line}}{\fmtlinenumber{ALG@line}}{}{}
\makeatother 
\let\fmtlinenumber\arabic 
\newcommand\mathspecalg[1]{\arabic{#1}'} 

\usepackage[colorlinks=true,linkcolor=blue,filecolor=blue,urlcolor=blue,citecolor=blue,pdftex,plainpages=false]{hyperref}

\journal{Computer Physics Communications}

\begin{document}

\begin{frontmatter}



\title{Efficient integration of gradient flow in lattice gauge theory and\\
properties of low-storage commutator-free
Lie group methods}


\author[ldep]{Alexei Bazavov}
\ead{bazavov@msu.edu}
\author[ldep]{Thomas Chuna}
\address[ldep]{Department of Computational Mathematics, Science and Engineering and
Department of Physics and Astronomy,\\
Michigan State University, East Lansing, MI 48824, USA}

\begin{abstract}
The smoothing procedure
known as the gradient flow
that suppresses ultraviolet fluctuations of gauge fields
plays an important role in lattice gauge theory calculations.
In particular, this procedure is often used for high-precision scale setting
and renormalization of operators. The gradient flow equation is defined on the
$\mbox{\textit{SU}}(3)$ manifold and therefore requires geometric, or structure-preserving,
integration methods to obtain its numerical solutions. We examine the properties and
origins of the three-stage third-order explicit Runge-Kutta Lie group integrator
commonly used in the lattice gauge theory community, demonstrate its relation to
$2N$-storage classical Runge-Kutta methods and
explore how its coefficients can be tuned
for optimal performance in integrating the gradient flow. We also compare the performance
of the tuned method with two third-order variable step size methods.
Next, based on the recently established connection between low-storage Lie group
integrators and classical $2N$-storage Runge-Kutta methods, we study two
fourth-order low-storage methods that provide a computationally efficient alternative
to the commonly used third-order method while retaining the convenient iterative
property of the latter. Finally, we demonstrate that almost no coding effort is
needed to implement the low-storage Lie group methods into existing gradient
flow codes.
\end{abstract}




\end{frontmatter}


\section{Introduction}

In lattice gauge theory~\cite{Wilson:1974sk},
a numerical approach to quantum gauge
theories, the path integrals are evaluated by sampling the space
of possible field configurations with a Markov Chain Monte Carlo
process and averaging over the Monte Carlo time series.
The fields are defined on a four-dimensional Euclidean space-time
grid and for the physically relevant case of Quantum Chromodynamics
(QCD) take values in the $\mbox{\textit{SU}}(3)$ group.

A smoothing procedure, referred to as \textit{gradient flow}, 
introduced by L\"{u}scher in Ref.~\cite{Luscher:2010iy},
allows one to evolve a given
gauge field configuration towards the classical solution. The gradient
flow possesses renormalizing properties and is often used for
renormalization of operators and determining the lattice spacing in
physical units.
Certain lattice calculations require determination of the lattice scale to
sub-percent precision.
Numerically, gradient flow amounts to integrating a
first-order differential equation on the $\mbox{\textit{SU}}(3)$ manifold. While a
variety of structure-preserving
integration methods can be used for this task~\cite{Hairer2006},
the three-stage third-order explicit Runge-Kutta integrator introduced
in Ref.~\cite{Luscher:2010iy} became the most commonly used method
in lattice gauge theory applications.

In this paper we explore the recently observed relations between classical
low-storage explicit Runge-Kutta methods and the commutator-free
Lie group methods~\cite{Bazavov2020}. We show how a three-stage
third-order integrator can be optimized specifically for integrating the
gradient flow and how higher-order methods with similar low-storage
properties can be constructed.

The paper is organized as follows. In Sec.~\ref{sec_lgt} we introduce the formalism
of lattice gauge theory and gradient flow. In Sec.~\ref{sec_rk} we review the 
properties of standard explicit Runge-Kutta integration methods
and so-called low-storage Runge-Kutta methods. In Sec.~\ref{sec_lie_int} we
discuss structure-preserving integrators, often called geometric
integrators or Lie group integrators. In Sec.~\ref{sec_num_res} we present
the numerical results on integrating the gradient flow on three lattice
ensembles with several third- and fourth-order Lie group methods.
We present our conclusions and recommendations for tuning the methods and
improving the computational efficiency of integrating the gradient flow
in Sec.~\ref{sec_concl}.

\section{Lattice gauge theory and the gradient flow}
\label{sec_lgt}
In lattice gauge theory
the primary degrees of freedom are $\mbox{\textit{SU}}(3)$ matrices $U_{x,\,\mu}$
that reside on the links of a hypercubic lattice with dimensions
$N_\sigma^3\times N_\tau$. The lattice spacing is $a$,
$x$ is an integer-valued four-vector, and $\mu=1,\dots,4$.
Gauge-invariant observables are
represented as traces of products of the gauge link variables along paths on the
lattice that are closed loops. The main observables we discuss here are the 
plaquette (4-link loop):
\begin{equation}
P_{x,\,\mu,\nu}=U_{x,\,\mu}U_{x+\mu,\,\nu}U^\dagger_{x+\nu,\,\mu}U^\dagger_{x,\,\nu}
\end{equation} 
rectangle (6-link loop):
\begin{equation}
R_{x,\,\mu,\nu}=U_{x,\,\mu}U_{x+\mu,\,\mu}U_{x+2\mu,\,\nu}
U^\dagger_{x+\mu+\nu,\,\mu}U^\dagger_{x+\nu,\,\mu}U^\dagger_{x,\,\nu}
\end{equation} 
and the so called clover expression constructed as a linear
combination of several plaquettes forming the shape of a clover
leaf:
\begin{equation}
C_{x,\,\mu,\nu}=\frac{i}{8}\left(Q_{x,\,\nu,\,\mu}-Q_{x,\,\mu,\nu}\right)
\end{equation}
where $Q_{x,\,\mu,\nu}=P_{x,\,\mu,\nu}+P_{x,\nu,-\mu}+P_{x,-\mu,-\nu}+P_{x,-\nu,\,\mu}$.

The simplest gauge action is the Wilson action~\cite{Wilson:1974sk}
that includes only the
plaquette term (for convenience we drop the factor
$1/g_0^2$ in the definition):
\begin{equation}
\label{eq_Swil}
S_{Wilson}=2\sum_x\sum_{\mu<\nu}{\rm Re}{\rm Tr}(1-P_{x,\,\mu,\nu}).
\end{equation} 
To suppress lattice discretization effects one can construct improved
actions such as, for instance, the tree-level Symanzik-improved gauge
action that includes the plaquette and rectangle terms~\cite{Luscher:1985zq}:
\begin{equation}
\label{eq_Ssym}
S_{S\!ymanzik}=\frac{5}{3}S_{Wilson}-\frac{1}{6}\sum_x\sum_{\mu\neq\nu}
{\rm Re}{\rm Tr}(1-R_{x,\,\mu,\nu}).
\end{equation} 
The clover action is
\begin{equation}
\label{eq_Scl}
S_{clover}=\frac{1}{2}\sum_x\sum_{\mu\neq\nu}{\rm Re}{\rm Tr}
(C_{x,\,\mu,\nu}C_{x,\,\mu,\nu})
\end{equation} 
and another variant of an improved action can be constructed as a linear
combination of the plaquette and clover terms.

To smoothen the fields and suppress ultraviolet fluctuations
Ref.~\cite{Luscher:2010iy} suggested evolving the gauge fields 
$U_{x,\,\mu}$ with the following gradient flow equation:
\begin{equation}
\label{eq_flow}
\frac{dV_{x,\,\mu}}{dt}=-\left\{\partial_{x,\,\mu}S^f(t)\right\}V_{x,\,\mu},
\,\,\,\,\,\,V_{x,\,\mu}(t=0)=U_{x,\,\mu}
\end{equation} 
where the differential operator $\partial_{x,\,\mu}$ acts on a function
of $\mbox{\textit{SU}}(3)$ group elements as defined in
Ref.~\cite{Luscher:2010iy} and
$S^f$ is the lattice action evaluated using the evolved gauge link variables $V(t)$.
We refer to the gradient flow as the
Wilson flow when $S^f=S_{Wilson}$ is used in the flow equation, and
as the Symanzik flow when $S^f=S_{S\!ymanzik}$. The flow time $t$ has dimensions
of lattice spacing squared.

One of the widespread applications of gradient flow in lattice gauge theory
is scale setting, \textit{i.e.} determination of the lattice spacing in
physical units for a given lattice ensemble. In this case the flow is run
until the flow time $t=w_0^2$~\cite{Borsanyi:2012zs} at which
\begin{equation}
\label{eq_w0}
\left[t\frac{d}{dt}t^2\langle S^o(t)\rangle\right]_{t=w_0^2}=Const
\end{equation}
and typically $Const=0.3$ is chosen. The lattice spacing is then set
by using the value of the $w_0$-scale in physical units, $w^{phys}_0$.
Eq.~(\ref{eq_w0}) is an improved version of the original proposal where
the action itself rather than its derivative was used~\cite{Luscher:2010iy}:
\begin{equation}
\label{eq_t0}
\left.t^2\langle S^o(t)\rangle\right|_{t=t_0}=Const.
\end{equation}
The observable used for the scale setting in Eq.~(\ref{eq_w0}) is
the action density $S^o$, not necessarily the same as $S^f$ in the flow
equation~(\ref{eq_flow}). As has been discussed in Ref.~\cite{Fodor:2014cpa}
different combinations of the flow action and the observable result in different
dependence on the lattice spacing.
Here we consider $S_{Wilson}$ and $S_{S\!ymanzik}$ for the flow
and $S_{Wilson}$, $S_{S\!ymanzik}$ and $S_{clover}$ for the observable.

\section{Classical Runge-Kutta methods}
\label{sec_rk}
Consider a first-order differential equation for a function $y(t)$
\begin{equation}
\label{eq_dydt}
\frac{dy}{dt}=f(t,y),
\end{equation}
and the initial condition $y(t=0)$ given.
An $s$-stage explicit Runge-Kutta (RK) method that propagates the numerical
approximation to the solution $y_t$ of Eq.~(\ref{eq_dydt})
at time $t$ to time $t+h$
is given in Algorithm~\ref{alg_class_RK}~\cite{ButcherBook,HairerBook1}.
\begin{algorithm}[h]
	\caption{Explicit classical $s$-stage Runge-Kutta method}
	\label{alg_class_RK}
	\begin{algorithmic}[1]
		\For{i=1,\dots,s}
		\State $y_i=y_t+h\sum_{j=1}^{i-1}a_{ij}k_j$\Comment{$a_{i,j\geqslant i}=0$}
		\State $k_i=f(t+hc_i,y_i)$\Comment{$c_1=0$}
		\EndFor
		\State $y_{t+h}=y_t+h\sum_{i=1}^{s}b_ik_i$
	\end{algorithmic}
\end{algorithm}

The self-consistency conditions require
\begin{equation}
c_i=\sum_{j=1}^{i-1}a_{ij}.
\end{equation}
We refer to this method as \textit{classical} RK method to distinguish it
from the Lie group integrators discussed in Sec.~\ref{sec_lie_int}.
To provide an order of accuracy $p$ the coefficients $a_{ij}$, $b_i$ need
to satisfy \textit{the order conditions}. The order conditions for
a classical RK method of third-order global accuracy are given in
\ref{sec_app_oc}. It is convenient to represent the set of
coefficients $a_{ij}$, $b_i$, $c_i$ as a \textit{Butcher tableau}, for instance,
for a 3-stage method (the first entry with $c_1=0$ is omitted):
\begin{equation}
\begin{array}
{c|lll}
c_2 & a_{21}\\
c_3 & a_{31} & a_{32} \\
\hline
& b_{1}   & b_{2}  & b_{3}  \\
\end{array}
\label{eq_abctable}
\end{equation}
The \textit{nodes} $c_i$ in the left column describe the time points at which
the stages are evaluated, the $a_{ij}$ in the middle give the \textit{weights}
of the right hand side function for each stage, and the bottom row gives the
weights for the final stage of the method. If the method is explicit each
stage can only depend on the previous ones and therefore the Butcher tableau
has a characteristic triangular shape.

For a 3-stage third-order classical RK method 
there are four order conditions and six independent
coefficients, thus, these methods belong to a two-parameter family.
Often, the coefficients $c_2$, $c_3$ are chosen as free parameters
and the rest are expressed through them, as given in
Eqs.~(\ref{eq_b2})--(\ref{eq_a31}).

In the following the discussion is restricted to autonomous problems where
the right hand side of Eq.~(\ref{eq_dydt}) does not explicitly depend on
time. Extension to non-autonomous problems is trivial.

\subsection{$2N$-storage classical RK methods}
\label{sec_2N_RK}

As is clear from Algorithm~\ref{alg_class_RK}, to compute
$y_{t+h}$ at the final step one needs to store $k_i$, $i=1,\dots,s$
(the right hand side evaluations) from all $s$ stages of the method.
It was shown in Ref.~\cite{WILLIAMSON198048} that a classical
RK method may be written in a form where only the values from the
previous stage are used. Therefore only two quantities need to
be stored at all times, independent of the number of stages of the method.
RK methods with such a property are called \textit{low-storage methods}.
A number of different types of low-storage methods have been developed
in the literature, \textit{e.g.}, 
Refs.~\cite{KENNEDY2000177,KETCHESON20101763}.
For the later discussion of Lie group
methods in Sec.~\ref{sec_lie_int} we focus on the methods of
Ref.~\cite{WILLIAMSON198048}, which are also called $2N$-storage
methods\footnote{Unlike other types of low-storage methods
(\textit{e.g.}, $2R$-, $2S$-, $3R$-, $3S$-, etc.) the $2N$-storage
methods have special properties that turned out to be related to Lie group
integrators~\cite{Bazavov2020}.}.

Given an $s$-stage RK method one can express its coefficients
through another set of coefficients $A_i$, $B_i$, $i=1,\dots,s$ such that
\begin{eqnarray}
a_{ij} &=& \left\{
\begin{array}{ll}
A_{j+1}a_{i,j+1}+B_j, & j<i-1,\\
B_j, & j=i-1,\\
0, & \mbox{otherwise},
\end{array}
\right.\label{eq_Ai}\\
b_i &=&\left\{
\begin{array}{ll}
A_{i+1}b_{i+1}+B_i, & i<s,\\
B_i, & i=s,
\end{array}
\right.\label{eq_Bi}
\end{eqnarray}
and for explicit methods necessarily $A_1=0$. A $2N$-storage
$s$-stage explicit classical RK method is given in
Algorithm~\ref{alg_class_2N_RK}.

\begin{algorithm}[h]
	\caption{$2N$-storage explicit classical $s$-stage Runge-Kutta method}
	\label{alg_class_2N_RK}
	\begin{algorithmic}[1]
		\State $y_0=y_t$
		\For{i=1,\dots,s}
		\State $\Delta y_i = A_i\Delta y_{i-1} + hf(y_{i-1})$\Comment{$A_1=0$}
		\State $y_i=y_{i-1}+B_i\Delta y_i$
		\EndFor
		\State $y_{t+h}=y_s$
	\end{algorithmic}
\end{algorithm}

\begin{figure}[h]
	\centering
	\includegraphics[width=0.8\columnwidth]{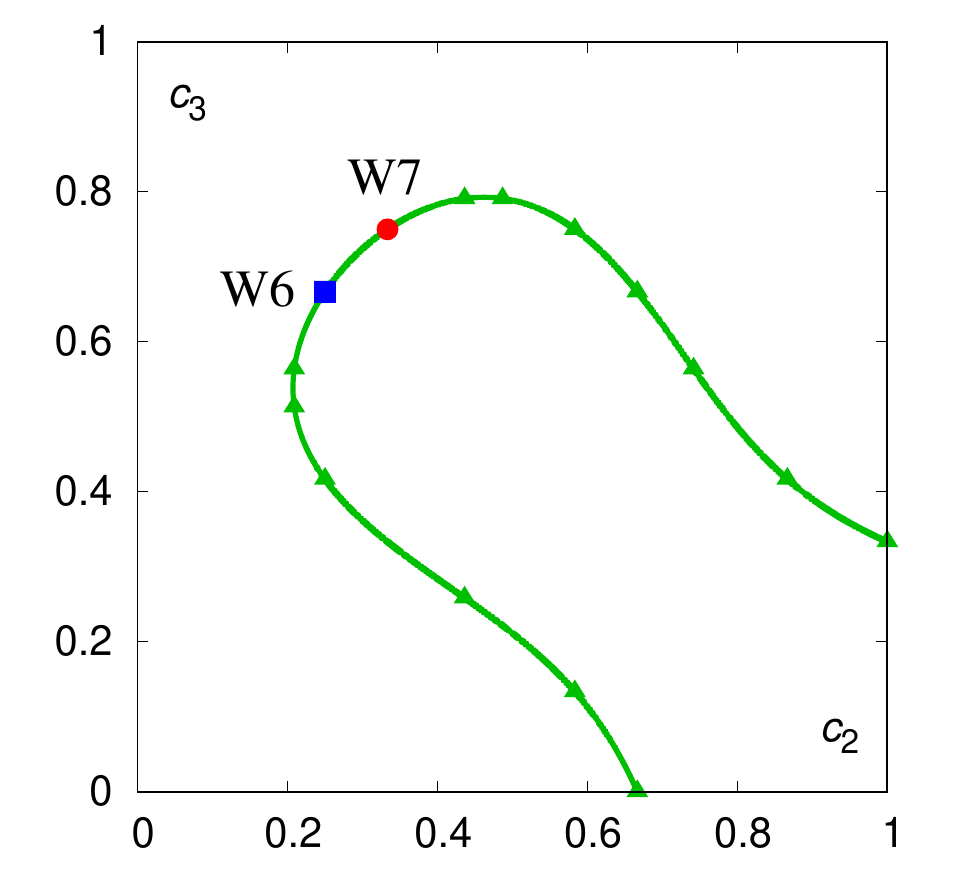}
	\caption{
		The Williamson curve, \textit{i.e.} the set of values of
			$c_2$ and $c_3$ coefficients for which the 3-stage third-order
			classical RK schemes can be written in the $2N$-storage format.
			The symbols (triangles, box and circle)
			correspond to rational solutions. The blue box and
			the red circle are the schemes that are discussed in more detail
			later. W6 and W7 labels indicate the original numbering in
			Ref.~\cite{WILLIAMSON198048}.
			There is a reflection symmetry along the $c_2+c_3=1$ line.
		\label{fig_w_rat}
	}
\end{figure}

For a 3-stage third-order method expressing the original $a_{ij}$, $b_i$
coefficients through $A_i$, $B_i$
leads to an additional, fifth, order condition for $a_{ij}$, $b_i$
that was found in~\cite{WILLIAMSON198048}. This means that the
coefficients of a $2N$-storage scheme now form a one-parameter family.
One can express the fifth order condition as an implicit function of
$c_2$ and $c_3$~\cite{WILLIAMSON198048}:
\begin{equation}
\label{eq_Wc2c3}
c_3^2(1-c_2)+c_3\left(c_2^2+\frac{1}{2}c_2-1\right)+
\left(\frac{1}{3}-\frac{1}{2}c_2\right)=0.
\end{equation}
This implicit function is shown in Fig.~\ref{fig_w_rat}.

While almost any point in the plane (except $c_2=c_3=1/3$) corresponds to a
possible 3-stage third-order classical RK coefficient scheme,
only the values on the curve correspond to classical RK methods that
can be rewritten in the $2N$-storage format, 
Algorithm~\ref{alg_class_2N_RK}. The plot shown in
Fig.~\ref{fig_w_rat} first appeared in~\cite{WILLIAMSON198048}
therefore we refer to it as the \textit{Williamson curve}.
To find coefficients for a 3-stage third-order $2N$-storage scheme
one can proceed in the following way:
\begin{itemize}
	\item Pick a value of $c_2$ in the allowed range.
	\item Solve Eq.~(\ref{eq_Wc2c3}) for $c_3$ and pick one of the branches.
	\item Express all $a_{ij}$, $b_i$ coefficients in terms of $c_2$ and $c_3$
	using Eqs.~(\ref{eq_b2})--(\ref{eq_a31}).
	\item Find $A_i$, $B_i$ by inverting the relations given
	in Eqs.~(\ref{eq_Ai}), (\ref{eq_Bi}).
\end{itemize}
The symbols on the Williamson curve are the points where $c_2$ and
$c_3$ (and all the other coefficients) have rational values.
These values are summarized in Table~\ref{tab_rat_w} in \ref{sec_app_rw}.
The schemes labeled with the blue box and red circle in the figure
play special role in our discussion later. We denote them:
\begin{itemize}
	\item RK3W6: $c_2=1/4$, $c_3=2/3$,
	\item RK3W7: $c_2=1/3$, $c_3=3/4$,
\end{itemize}
where ``RK3'' means that the method is of the third order of global
accuracy and ``W6'' and ``W7'' preserve the numbering used
in~\cite{WILLIAMSON198048} where these $2N$-storage schemes
first appeared.

For $2N$-storage schemes with more than three stages and orders
higher than three there are no analytic solutions available.
The coefficients can be found by expressing the order conditions
through the coefficients $A_i$, $B_i$ and solving
the resulting system of non-linear equations numerically.
Multiple $2N$-storage schemes have been
designed in this way in the
literature~\cite{CK1994,BERNARDINI20094182,BERLAND20061459,TOULORGE20122067,STANESCU1998674,ALLAMPALLI20093837,NIEGEMANN2012364,Yan2017}.

\subsection{Variable step size methods}

In the previous sections we considered integration methods that
operate with fixed step size. If an estimate of the local error is
available one can adjust the step size during the integration. Methods with
such a property are known as \textit{variable step size} or
\textit{adaptive} integrators. To construct a variable step size method
one uses two schemes of different order simultaneously. The difference
between the two solutions after one step of integration serves as an
estimate of the local error and is used to adjust the step size.

Later in Sec.~\ref{sec_order3_vs}
we will consider two variable step size schemes where a third-order
integrator is used for propagating the solution and an embedded
second-order integrator is used to construct an alternative estimate
of the solution. Let $d$ be some measure of distance between the solutions
(the precise metric used is not important at this point) after the current
integration step and $\delta$ a fixed parameter -- local tolerance.
After an integration step the step size is adjusted as
\begin{equation}
h\to 0.95\sqrt[3]{\frac{\delta}{d}}\,h.
\end{equation}
If $d>\delta$ then the integration step is redone with the adjusted $h$.
Otherwise the integration proceeds with the adjusted $h$. Such a procedure
ensures that the local error at every step is bounded by $\delta$.

It is important to note that setting a bound on the \textit{local} error does not
actually tell one what value of the \textit{global} error will be achieved. We
discuss this point in more detail in Sec.~\ref{sec_order3_vs}.

\section{Lie group integrators}
\label{sec_lie_int}
Consider now a differential equation on a manifold:
\begin{equation}
\label{eq_dYdt}
\frac{dY}{dt}=F(Y)Y.
\end{equation}
The results in this section are valid in general for $Y$ taking
values on an arbitrary manifold equipped with a group action.
For the main discussion that follows, $Y$ will represent a gauge link
variable,
which is an $\mbox{\textit{SU}}(3)$ matrix. Capital letters are used here to
emphasize that the variables do not necessarily
commute, unlike in the classical RK case.

If one uses a classical RK scheme for solving Eq.~(\ref{eq_dYdt})
numerically, an update of the form $Y+Const\cdot hF(Y)Y$ will 
move $Y$ away from the manifold. One needs to use
\textit{geometric, or structure-preserving} integration schemes that
update $Y$ as $\exp(Const\cdot hF(Y))Y$. We discuss the two
main approaches to constructing such methods next.

\subsection{Munthe-Kaas Lie group methods}

Let us first return to the classical RK Algorithm~\ref{alg_class_RK}.
One can modify the step 2 in the following way: Evaluate the sum
in the second term, exponentiate it and act on $y_t$, \textit{i.e.}
$Y_i=\exp(h\sum_{j=1}^{i-1}a_{ij}K_j)Y_t$, where $K_j=F(Y_j)$.
In this case, however, the extra uncanceled terms in the Taylor expansion
of the scheme will result in a method whose global order of accuracy is lower
than for the classical scheme. This can be cured by adding commutators.
A scheme given in
Algorithm~\ref{alg_RKMK} which is referred to as
the Runge-Kutta-Munthe-Kaas (RKMK) method was introduced in
Ref.~\cite{MUNTHEKAAS1999115}.
There, the expansion of the inverse
derivative of the matrix exponential $d\exp_{\cal U}^{-1}({\cal V})$ is truncated at
the order $p-1$ (to distinguish it from the full series it is commonly denoted
``dexpinv''):
\begin{equation}
\label{eq_dexpinv}
{\rm dexpinv}({\cal U},{\cal V},p)=\sum_{k=0}^{p-1}\frac{B_k}{k!}{\rm ad}_{\cal U}^k({\cal V}),
\end{equation}
$B_k$ are the Bernoulli numbers and the $k$-th power of the adjoint
operator ${\rm ad}_{\cal U}({\cal V})$ is given by an iterated commutator application:
\begin{eqnarray}
{\rm ad}_{\cal U}^0({\cal V})&=&{\cal V},\\
{\rm ad}_{\cal U}^1({\cal V})&=&[{\cal U},{\cal V}],\\
{\rm ad}_{\cal U}^k({\cal V})&=&{\rm ad}_{\cal U}({\rm ad}_{\cal U}^{k-1}({\cal V}))\nonumber\\
&=&[{\cal U},[{\cal U},[\dots,[{\cal U},{\cal V}]]]].
\end{eqnarray}
This algorithm results in a Lie group integrator of order $p$ whose
coefficients are the coefficients of a classical RK method of order $p$.
\begin{algorithm}[h]
	\caption{$s$-stage Runge-Kutta-Munthe-Kaas Lie group method}
	\label{alg_RKMK}
	\begin{algorithmic}[1]
		\For{i=1,\dots,s}
		\State ${\cal U}_i=h\sum_{j=1}^{i-1}a_{ij}\tilde K_j$\Comment{$a_{i,j\geqslant i}=0$}
		\State $Y_i=\exp({\cal U}_i)Y_t$
		\State $K_i=F(Y_i)$
		\State $\tilde K_i = {\rm dexpinv}({\cal U}_i,K_i,p)$
		\EndFor
		\State ${\cal V}=\sum_{i=1}^{s}b_i\tilde K_i$
		\State $Y_{t+h}=\exp({\cal V})Y_t$
	\end{algorithmic}
\end{algorithm}

The advantage of Algorithm~\ref{alg_RKMK} is that any classical RK
method can be turned into a Lie group integrator. The number of
commutators can often be reduced as discussed
in~\cite{MuntheKaasOwren1999}. It was found, however, that schemes
that avoid commutators can be more stable and provide lower global
error at the same computational cost. We discuss them next.

\subsection{Commutator-free Lie group methods}

The earlier work on manifold integrators~\cite{Crouch1993}
was extended in Ref.~\cite{CELLEDONI2003341}
to design a class of \textit{commutator-free}
Lie group methods where each stage of the method may include
a composition of several exponentials.
A general scheme is given
in Algorithm~\ref{alg_CF}. We use the notation of Ref.~\cite{Bazavov2020}
instead of the original notation of Ref.~\cite{CELLEDONI2003341}.
$L_i$ is the number of exponentials used at stage $i$, $J_{il}$ is
the number of right hand side evaluations $K_j$ used in the $l$-th
exponential at stage $i$, $L$ is the number of exponentials at the final
step, and $I_l$ is the number of right hand side evaluations $K_i$
used in the $l$-th exponential at the final step. ${\cal T}$ represents
a ``time-ordered'' product meaning that an exponential with a lower
value of index $l$ is located \textit{to the right}.

\begin{algorithm}[h]
	\caption{$s$-stage commutator-free Lie group method}
	\label{alg_CF}
	\begin{algorithmic}[1]
		\State $Y_1=Y_t$, $K_1=F(Y_1)$
		\For{i=2,\dots,s}
		\State $Y_i = {\cal T}\left\{\prod_{l=1}^{L_i}\exp\left(h \sum_{j=1}^{J_{il}} \alpha_{l;ij}K_j\right)\right\}Y_t$
		\State $K_i=F(Y_i)$
		\EndFor
		\State $Y_{t+h} = {\cal T}\left\{\prod_{l=1}^{L}
		\exp\left(h \sum_{i=1}^{I_l} \beta_{l;i}K_i\right)\right\}Y_t$
	\end{algorithmic}
\end{algorithm}

The coefficients $\alpha_{l;ij}$, $\beta_{l;i}$ are related to the
coefficients of a classical RK method as~\cite{CELLEDONI2003341}
\begin{equation}
\sum_{l=1}^{L_i}\alpha_{l;ij}=a_{ij},\,\,\,\,\,\,
\sum_{l=1}^{L}\beta_{l;i}=b_{i}.\label{eq_abCMO}
\end{equation}
To better understand the notation of Algorithm~\ref{alg_CF} we list
in Algorithm~\ref{alg_CF_ex}
explicit steps of
one of the methods of Ref.~\cite{CELLEDONI2003341} where
$s=3$, $L_1=0$, $L_2=1$, $J_{21}=1$, $L_3=1$, $J_{31}=2$, $L=2$,
$I_1=1$ and $I_2=3$
\begin{algorithm}[h]
	\caption{$3$-stage third-order commutator-free Lie group method
	of Ref.~\cite{CELLEDONI2003341}}
	\label{alg_CF_ex}
	\begin{algorithmic}[1]
		\State $Y_1=Y_t$
		\State $K_1=F(Y_1)$
		\State $Y_2=\exp(h\alpha_{1;21}K_1)Y_t$
		\State $K_2=F(Y_2)$
		\State $Y_3=\exp(h(\alpha_{1;32}K_2+\alpha_{1;31}K_1))Y_t$
		\State $K_3=F(Y_3)$
		\State $Y_{t+h} = \exp(h(\beta_{2;3}K_3+\beta_{2;2}K_2+\beta_{2;1}K_1))\exp(h\beta_{1;1}K_1)Y_t$
	\end{algorithmic}
\end{algorithm}

It was found in Ref.~\cite{CELLEDONI2003341} that fixing
$\beta_{1;1}=\alpha_{1;21}=1/3$ allows one to reuse $Y_2$ at the final
step and the other coefficients form a one-parameter family of solutions.

\subsection{Low-storage commutator-free Lie group methods}
\label{sec_ls_lie}

Ref.~\cite{CELLEDONI2003341} considered such commutator-free
methods that reuse exponentials. For instance, in Algorithm~\ref{alg_CF_ex}
$Y_2$ is reused at the final stage
so one needs only three exponential evaluations in total.
Recently, Ref.~\cite{Bazavov2020} considered designing a commutator-free
method where every next stage reuses $Y_i$ from the previous stage and
contains only one exponential evaluation per stage
(inspired by Algorithm~\ref{alg_W6} of Ref.~\cite{Luscher:2010iy}, see below).
Such methods form a subclass
of methods of Ref.~\cite{CELLEDONI2003341} but differ from the solutions
found there by how the exponentials are reused.
It turned out that for a 3-stage third-order commutator-free 
Lie group method with exponential reuse
the additional order condition resulting
from non-commutativity is the same as the order condition for
a $2N$-storage 3-stage third-order classical RK method, Eq.~(\ref{eq_Wc2c3}).
Thus, it was proven in~\cite{Bazavov2020} 
that all $2N$-storage 3-stage third-order classical RK methods
of~\cite{WILLIAMSON198048}, \textit{i.e.} all points on the Williamson
curve, Fig.~\ref{fig_w_rat}, are also low-storage third-order commutator-free
Lie group integrators. It was conjectured in~\cite{Bazavov2020}
that $2N$-storage classical RK methods of order higher than three are also
automatically Lie group integrators of the same order. Numerical
evidence was provided in support of the conjecture.
Moreover, for a given set of numerical values of coefficients
$A_i$, $B_i$ of a classical $2N$-storage RK method the order
of the Lie group method based on it can be determined
algorithmically by using B-series~\cite{Knoth2020git}.
Thus for all such methods that we use here, 
given the coefficients $A_i$, $B_i$ of a classical
$2N$-storage RK method with $s$ stages and global
order of accuracy $p$, the procedure listed in Algorithm~\ref{alg_CF_2N_RK}
is a low-storage commutator-free Lie group method of order $p$.

\begin{algorithm}[h]
	\caption{$2N$-storage $s$-stage 
		commutator-free Runge-Kutta Lie group method}
	\label{alg_CF_2N_RK}
	\begin{algorithmic}[1]
		\State $Y_0=Y_t$
		\For{i=1,\dots,s}
		\State $\Delta Y_i = A_i\Delta Y_{i-1} + hF(Y_{i-1})$\Comment{$A_1=0$}
		\State $Y_i=\exp(B_i\Delta Y_i)Y_{i-1}$
		\EndFor
		\State $Y_{t+h}=Y_s$
	\end{algorithmic}
\end{algorithm}

Let us now turn to the discussion of the integrator first introduced by L\"{u}scher
in Ref.~\cite{Luscher:2010iy}. In our notation it is given in
Algorithm~\ref{alg_W6}. This scheme belongs to the generic class of commutator-free
Lie group methods developed in Ref.~\cite{CELLEDONI2003341}, however, it differs
from the classes of solutions found there. Given that the linear combination
of $K_1$ and $K_2$ is the same at steps~\ref{a_s3} and \ref{a_sf} and the previous
stage is reused at steps~\ref{a_s2}, \ref{a_s3} and \ref{a_sf}, this integrator
has certain reusability property. As far as we are aware, an integrator with the
structure and numerical coefficients of Algorithm~\ref{alg_W6} was not present
in the literature on manifold integrators prior to Ref.~\cite{Luscher:2010iy}.
Thus, we believe that this method was derived independently. Since its derivation
was not presented in~\cite{Luscher:2010iy}, we present our derivation in
\ref{sec_app_der} for illustrative purposes and also to document the order
conditions in the form that we were not able to find in the existing
literature. This scheme provides a link to the recent developments
of Ref.~\cite{Bazavov2020}.

\begin{algorithm}[h]
	\caption{3-stage third-order Lie group method of Ref.~\cite{Luscher:2010iy}}
	\label{alg_W6}
	\begin{algorithmic}[1]
	\State $Y_1=Y_t$
	\State $K_1=F(Y_1)$
	\State $Y_2=\exp\left(h\frac{1}{4}K_1\right)Y_1$
	\label{a_s2}
	\State $K_2=F(Y_2)$
	\State $Y_3=\exp\left(h\left(\frac{8}{9}K_2-\frac{17}{36}K_1\right)\right)Y_2$
	\label{a_s3}
	\State $K_3=F(Y_3)$
	\State $Y_{t+h} = \exp\left(h\left(\frac{3}{4}K_3-\frac{8}{9}K_2+\frac{17}{36}K_1\right)\right)Y_3$
	\label{a_sf}
\end{algorithmic}
\end{algorithm}

It turns out that when Algorithm~\ref{alg_W6} is rewritten in the
$2N$-storage format of Algorithm~\ref{alg_CF_2N_RK} and the
$A_i$, $B_i$ coefficients are converted to the coefficients of the classical
underlying RK scheme, Eqs.~(\ref{eq_Ai}), (\ref{eq_Bi}),
the latter are the same
as for the RK3W6 classical RK method discussed in Sec.~\ref{sec_2N_RK}.
It is shown as a blue square on the Williamson curve, Fig.~\ref{fig_w_rat}. Thus,
the method of Ref.~\cite{Luscher:2010iy} belongs to the class of $2N$-storage
classical RK methods which are automatically Lie group integrators of the same
order as proven in~\cite{Bazavov2020}. We will explore this to find the optimal set
of coefficients for integrating the gradient flow in Sec.~\ref{sec_order3}.

\section{Numerical results}
\label{sec_num_res}
To explore the properties of different integrators we used three gauge ensembles
with the lattice spacing ranging from $0.15$~fm down to $0.09$~fm
listed in
Table~\ref{tab_ensembles}. These ensembles were generated by the MILC collaboration
with the one-loop improved gauge action~\cite{Luscher:1985zq}
and the Highly Improved Staggered Quark (HISQ)
action~\cite{Follana:2006rc,Bazavov:2010ru}. 
The light quark masses were tuned to produce the
Goldstone pion mass of about $300$~MeV and the strange and charm quark
masses are set to the physical values.

\begin{table}
	\small
	\centering
	\caption{
		The MILC 2+1+1 flavor ensembles used in this study,
		the details can be found in~\cite{Bazavov:2012xda}.
		In the second column the volume is listed,
		in the third the approximate lattice spacing and 
		in the fourth the maximum flow time $T_{max}$. Here $T_{max}$ is
		dimensionless and approximately equal to $(w_0^{phys}/a)^2$.
		\label{tab_ensembles}
	}
	\vspace{8pt}
	\begin{tabular}{lrrr}
		Ensemble & $N_\sigma^3\times N_\tau$ & $a$, fm & $T_{max}$  \\
		\hline
		\texttt{l1648f211b580m013m065m838}   & $16^3\times48$ & $0.15$ & $1.4$ \\
		\texttt{l2464f211b600m0102m0509m635} & $24^3\times64$ & $0.12$ & $2.0$ \\
		\texttt{l3296f211b630m0074m037m440}  & $32^3\times96$ & $0.09$ & $3.8$ \\
	\end{tabular}
\end{table}

We integrate the flow for the amount of time $T_{max}$ that is needed 
to determine the $w_0$-scale according to Eq.~(\ref{eq_w0})
with $Const=0.3$. The values
of $T_{max}$ for each ensemble are given in Table~\ref{tab_ensembles}.
The $w_0$-scale for these ensembles was determined in~\cite{Bazavov:2015yea}.

To find the global error for each integration method we need to compare to the
exact solution. For this purpose we have also implemented a 13-stage
eighth-order RK integrator of Munthe-Kaas type, Algorithm~\ref{alg_RKMK},
with Prince-Dormand coefficients~\cite{PRINCE198167}
that we refer to as RKMK8. At step size $h=10^{-2}$ it provides the result
that is exact within the floating point double precision.
For this reason the results from RKMK8 are labeled as ``exact.''

To evaluate the global error introduced by the integration methods
several quantities are studied.
Let us first define the squared distance between two $\mbox{\textit{SU}}(3)$ matrices
$X$, $Y$:
\begin{equation}
\label{eq_dist_sq}
{\cal D}(X,Y)\equiv\sum_{i,j=1}^3|X_{ij}-Y_{ij}|^2.
\end{equation}
For a set of flowed gauge fields
the distance from the exact solution is defined as
\footnote{Other definitions are possible, \textit{e.g.}
$$
\Delta V\equiv\sqrt{\sum_x\sum_\mu{\cal D}(V_{x,\,\mu}(t=T_{max}),V^{exact}_{x,\,\mu}(t=T_{max}))}.
$$
They scale with the step size in the same way as~(\ref{eq_dist_norm}).}:
\begin{equation}
\label{eq_dist_norm}
\Delta V\equiv\sum_x\sum_\mu\sqrt{{\cal D}(V_{x,\,\mu}(t=T_{max}),V^{exact}_{x,\,\mu}(t=T_{max}))}.
\end{equation}
We also calculate the value of the plaquette, rectangle and clover expression
averaged over the lattice:
\begin{eqnarray}
{\cal P}&=&\frac{1}{6N_\sigma^3N_\tau}\sum_x\sum_{\mu<\nu}{\rm Re}{\rm Tr}(P_{x,\,\mu\nu}),\label{eq_P}\\
{\cal R}&=&\frac{1}{12N_\sigma^3N_\tau}\sum_x\sum_{\mu\neq\nu}{\rm Re}{\rm Tr}(R_{x,\,\mu\nu}),\label{eq_R}\\
{\cal C}&=&\frac{1}{N_\sigma^3N_\tau}\sum_x\sum_{\mu\neq\nu}{\rm Re}{\rm Tr}(C_{x,\,\mu\nu}C_{x,\,\mu\nu}).\label{eq_C}
\end{eqnarray}
Up to a constant prefactor and a shift these quantities provide the three
different discretizations of the action,
Eqs.~(\ref{eq_Swil})--(\ref{eq_Scl}) that can be used
for the $w_0$-scale determination in~(\ref{eq_w0}). The global
error is also estimated from the differences:
\begin{eqnarray}
\label{eq_dZ}
\Delta Z\equiv Z(t)-Z^{exact}(t),
\end{eqnarray}
where $Z={\cal P}$, ${\cal R}$ or ${\cal C}$ and $t$-dependence means that
these quantities are computed using evolved gauge links $V(t)$.

\subsection{Tuning the third-order low-storage Lie group integrator}
\label{sec_order3}

We now address the question of what integrator out of the family of schemes
along the Williamson curve can provide the lowest error for integrating
the $\mbox{\textit{SU}}(3)$ gradient flow. The Williamson curve is parametrized with a variable
$u$ which is the distance along the curve from the point $(2/3,0)$ to $(1,1/3)$
such that $u\in [0,1]$. We pick 32 coefficient schemes $c_2(u)$, $c_3(u)$
and use their coefficients for low-storage commutator-free Lie group integrators
in the form of Algorithm~\ref{alg_CF_2N_RK}.
We refer to these methods in general as LSCFRK3 -- 
low-storage, commutator-free, Runge-Kutta, third-order.
We picked such values of $c_2$ and $c_3$ that are either
rational or given in terms of radicals.
Two particular schemes are of
interest in the following: LSCFRK3W6 (equivalent to the integrator of
Ref.~\cite{Luscher:2010iy}) and LSCFRK3W7. These are 
commutator-free Lie group versions of
the classical RK integration schemes RK3W6 and RK3W7 discussed in 
Sec.~\ref{sec_2N_RK}.

For this part of the calculation we have chosen 11 lattices per each ensemble
separated by 500 molecular dynamics time units for the first two ensembles and
360 time units for the third ensemble in Table~\ref{tab_ensembles}.
For the first two ensembles we ran both the Wilson and Symanzik flow, and
only the Symanzik flow for the third ensemble.
We ran all 32 LSCFRK3 methods at step sizes $h=1/16$, $1/32$, $1/64$ and $1/128$
and RKMK8 at $h=1/128$.

\begin{figure}[h]
	\centering
	\includegraphics[width=\columnwidth]{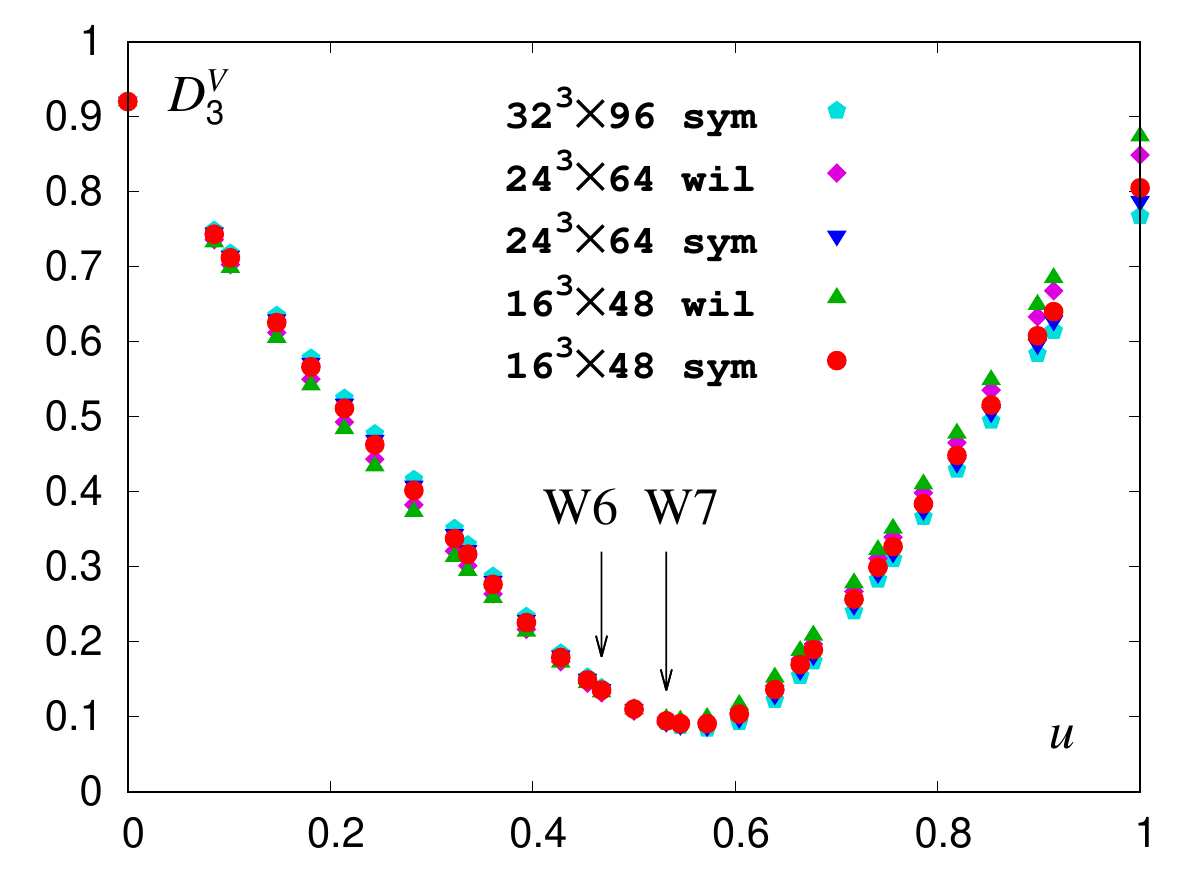}
	\caption{
		The leading-order coefficient $D_3^V$ for the global integration error defined
		in Eq.~(\ref{eq_DV_fit}) as function of the distance along the
		Williamson curve $u$. $u=0$ represents the LSCFRK3 method with
		$c_2=2/3$, $c_3=0$, and $u=1$ the method with $c_2=1$, $c_3=1/3$.
		The arrows labeled ``W6'' and ``W7'' represent the LSCFRK3W6 and
		LSCFRK3W7 schemes discussed in the text. The statistical errors
		are (much) smaller than the symbol size.
		\label{fig_norm3}
	}
\end{figure}
We first consider the behavior of the distance metric defined 
in~(\ref{eq_dist_norm}). For a third-order method the distance
is expected to scale as $O(h^3)$. We fit the distance $\Delta V$ as a function of the
step size to a polynomial form:
\begin{equation}
\label{eq_DV_fit}
\langle\Delta V(h)\rangle=D_3^V h^3+D_4^V h^4+D_5^V h^5,
\end{equation}
where $\langle\dots\rangle$ represents the ensemble average.
In some cases we omitted the fifth-order term if a reasonable fit resulted from
just the first two terms. The errors on the fit parameters were estimated with
a single elimination jackknife procedure.

In Fig.~\ref{fig_norm3} we show the dependence of the leading-order global
error coefficient $D_3^V$ as function of the distance $u$ along the Williamson curve
(\textit{i.e.} the RK coefficient scheme) for all ensembles and flows that
we analyzed. The values of $D_3^V$ are normalized in the following way.
For the Symanzik flow on the $16^3\times48$ lattice $D_3^V(u)$ is divided
by $5\times10^6$. (This large factor stems from the fact that our definition
of $\Delta V$ is extensive.) For all the other ensembles and flows $D_3^V(u)$
is divided by such a constant that $D_3^V(u=0)$ coincides with the one for the
$16^3\times48$ lattice Symanzik flow. As one can observe from Fig.~\ref{fig_norm3},
the behavior of $D_3^V(u)$ is similar for different ensembles and different types
of flow. The LSCFRK3W7 Lie group integrator is closer to the minimum of the global
error than the LSCFRK3W6 method. Note that since the definition of $\Delta V$
is manifestly positive, the leading order coefficient $D_3^V(u)$ is also positive
for all $u$.

\begin{figure}[h]
	\centering
	\includegraphics[width=\columnwidth]{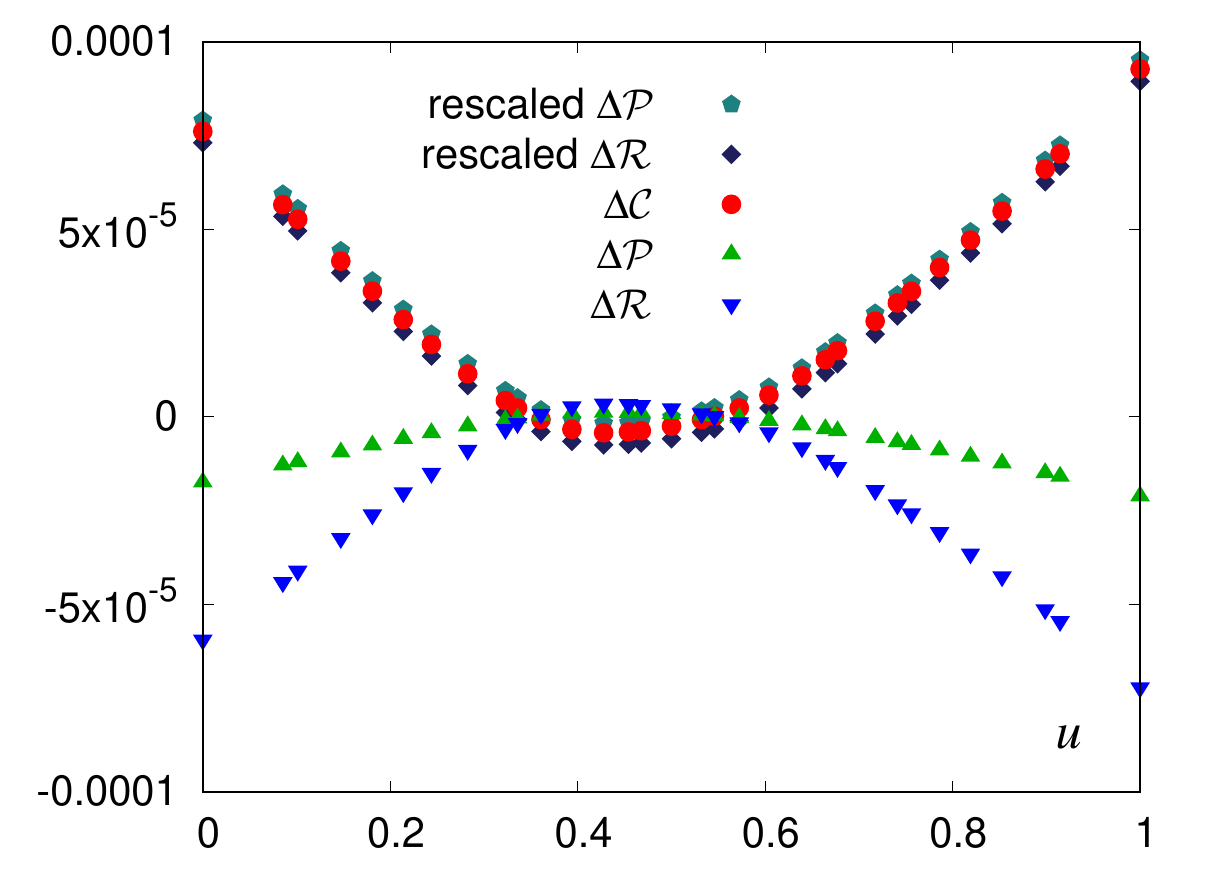}
	\caption{
		The global integration error in the plaquette
		$\langle\Delta{\cal P}\rangle$,
		rectangle $\langle\Delta{\cal R}\rangle$
		and the clover expression $\langle\Delta{\cal C}\rangle$
		as function of distance along the Williamson curve $u$.
		$\langle\Delta{\cal P}\rangle$ and $\langle\Delta{\cal R}\rangle$
		match $\langle\Delta{\cal C}\rangle$ after
		they are multiplied by a constant factor. The rescaled values of
		$\langle\Delta{\cal P}\rangle$ and $\langle\Delta{\cal R}\rangle$
		are shifted by $\pm3\times10^{-6}$
		to be distinguishable from $\langle\Delta{\cal C}\rangle$.
		\label{fig_prc}
	}
\end{figure}
Next, we study the action related observables, Eqs.~(\ref{eq_P})--(\ref{eq_C}).
In Fig.~\ref{fig_prc} we show the global errors $\langle\Delta{\cal P}\rangle$, 
$\langle\Delta{\cal R}\rangle$ and $\langle\Delta{\cal C}\rangle$,
defined in~(\ref{eq_dZ}), as function of $u$ evaluated
at $T_{max}=1.4$ with step size $h=1/16$ for the Wilson flow on the $a=0.15$~fm ensemble.
When $\langle\Delta{\cal P}\rangle$ and $\langle\Delta{\cal R}\rangle$
are rescaled by a constant so that they
coincide with $\langle\Delta{\cal C}\rangle$ at $u=0$, all three quantities collapse onto the
same curve. In fact, the collapse is so accurate that the quantities labeled ``rescaled''
in the figure would be completely covered by the clover
$\langle\Delta{\cal C}\rangle$ had not
we shifted them by $\pm 3\times10^{-6}$. This is expected since at later
flow times the gauge fields are smooth and the difference between different
discretizations diminishes. It is remarkable that the collapse happens already at our
coarsest lattice, for the least improved flow at the largest step size, $h=1/16$.
We therefore focus solely on the clover discretization in the following.

Similarly to Eq.~(\ref{eq_DV_fit}) we fit the global integration error
$\Delta {\cal C}$ as function of step size to a polynomial:
\begin{equation}
\label{eq_DC_fit}
\langle\Delta {\cal C}(h)\rangle=D_3^C h^3+D_4^C h^4+D_5^C h^5.
\end{equation}
We omit the fifth-order term whenever a reasonable fit is obtained with the first two
terms.

\begin{figure}[h]
	\centering
	\includegraphics[width=\columnwidth]{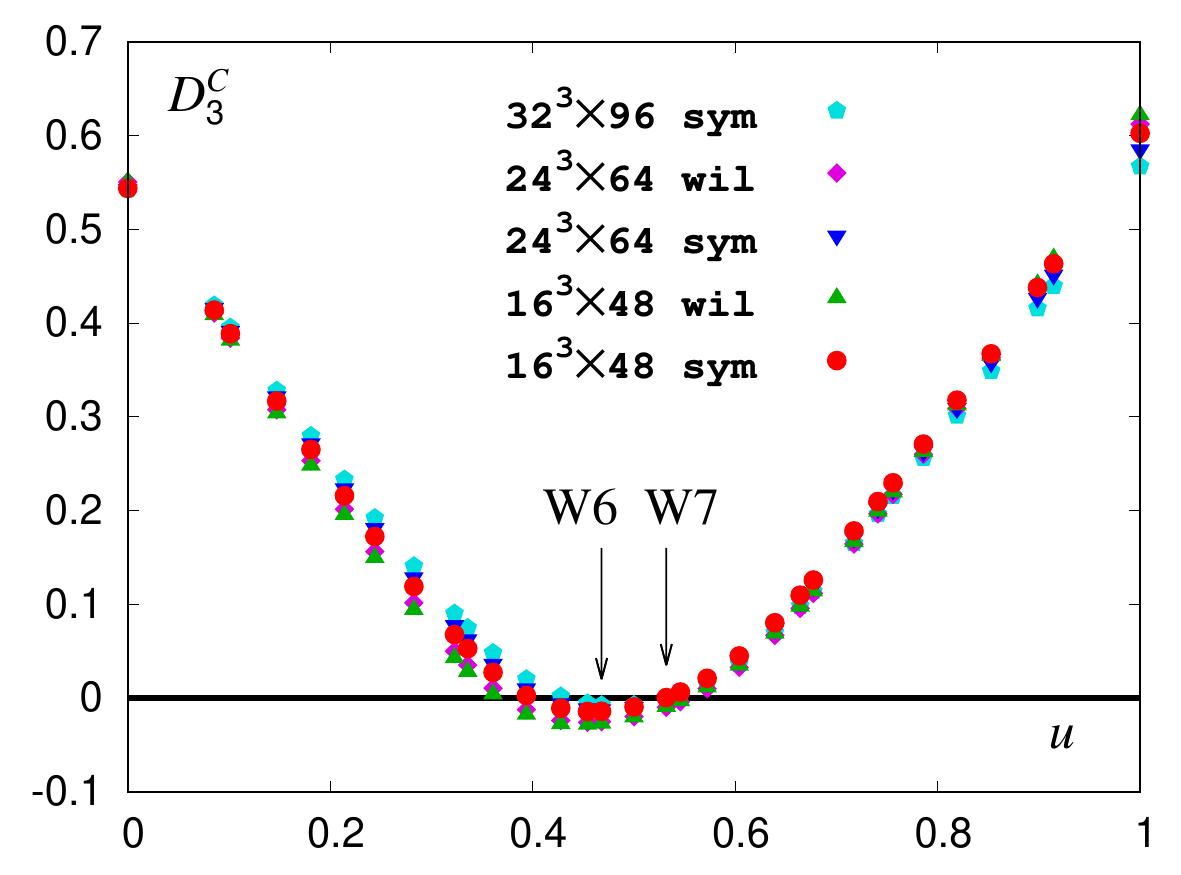}
	\caption{
		The leading-order coefficient $D_3^C$ for the global integration error defined
		in Eq.~(\ref{eq_DC_fit}) as function of the distance along the
		Williamson curve $u$. 
		The arrows labeled ``W6'' and ``W7'' represent the LSCFRK3W6 and
		LSCFRK3W7 schemes discussed in the text. The statistical errors
		are (much) smaller than the symbol size.
		\label{fig_clover3}
	}
\end{figure}
In Fig.~\ref{fig_clover3} we plot the dependence of the leading order
global error coefficient $D_3^C$ as function of $u$ for all ensembles and flows,
similar to Fig.~\ref{fig_norm3}. The data is rescaled 
such that all values at $u=0$ match the value of $D_3^C$ for the Symanzik flow
on the $16^3\times48$ lattice. 
Interestingly, unlike $\langle \Delta V(h)\rangle$, 
$\langle \Delta {\cal C}(h)\rangle$ is not necessarily positive. We observe that while most 
of the integration schemes approach the exact result from above, there is a region
of $u$ where the exact result is approached from below. The LSCFRK3W7 scheme
is close to the point where $D_3^C=0$ universally across the ensembles and types
of flow.

\begin{figure}[h]
	\centering
	\includegraphics[width=\columnwidth]{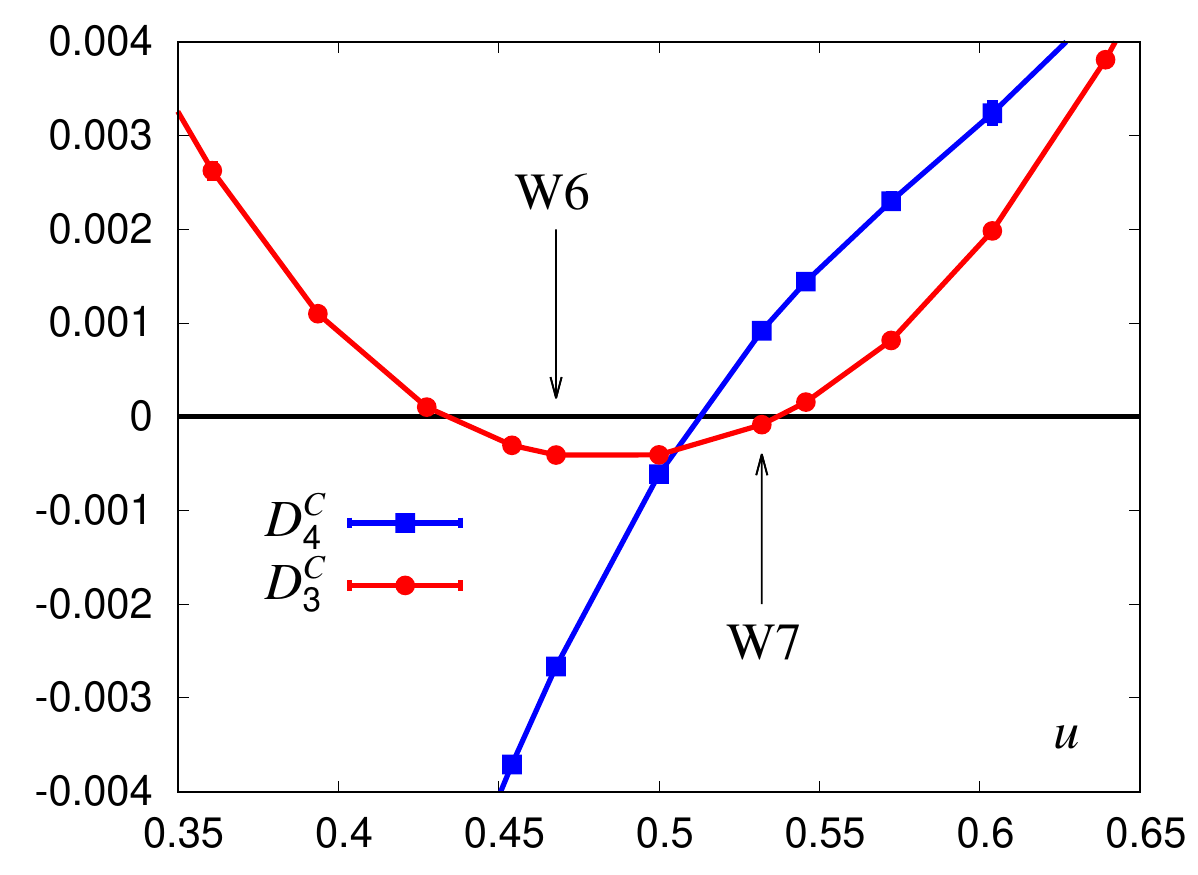}
	\caption{
		The leading $D_3^C$ and the next-to-leading order $D_4^C$ coefficients
		for the global integration error defined
		in Eq.~(\ref{eq_DC_fit}) for the Symanzik flow on $32^3\times96$
		ensemble. The lines are not fits but are drawn to guide the eye.
		The $D_4^C$ coefficient
		is divided by a factor of two to better fit in the frame.
		\label{fig_clover_zoom}
	}
\end{figure}
The interval $u\in[0.35,0.65]$ is magnified in Fig.~\ref{fig_clover_zoom} for
the $32^3\times96$ ensemble and the Symanzik flow.
The fourth-order coefficient
$D_4^C(u)$ is also shown. For the LSCFRK3W7 scheme it is also relatively small.
Therefore this method provides close to the lowest error for the action
observables that are central for scale setting.

\begin{figure}[h]
	\centering
	\includegraphics[width=\columnwidth]{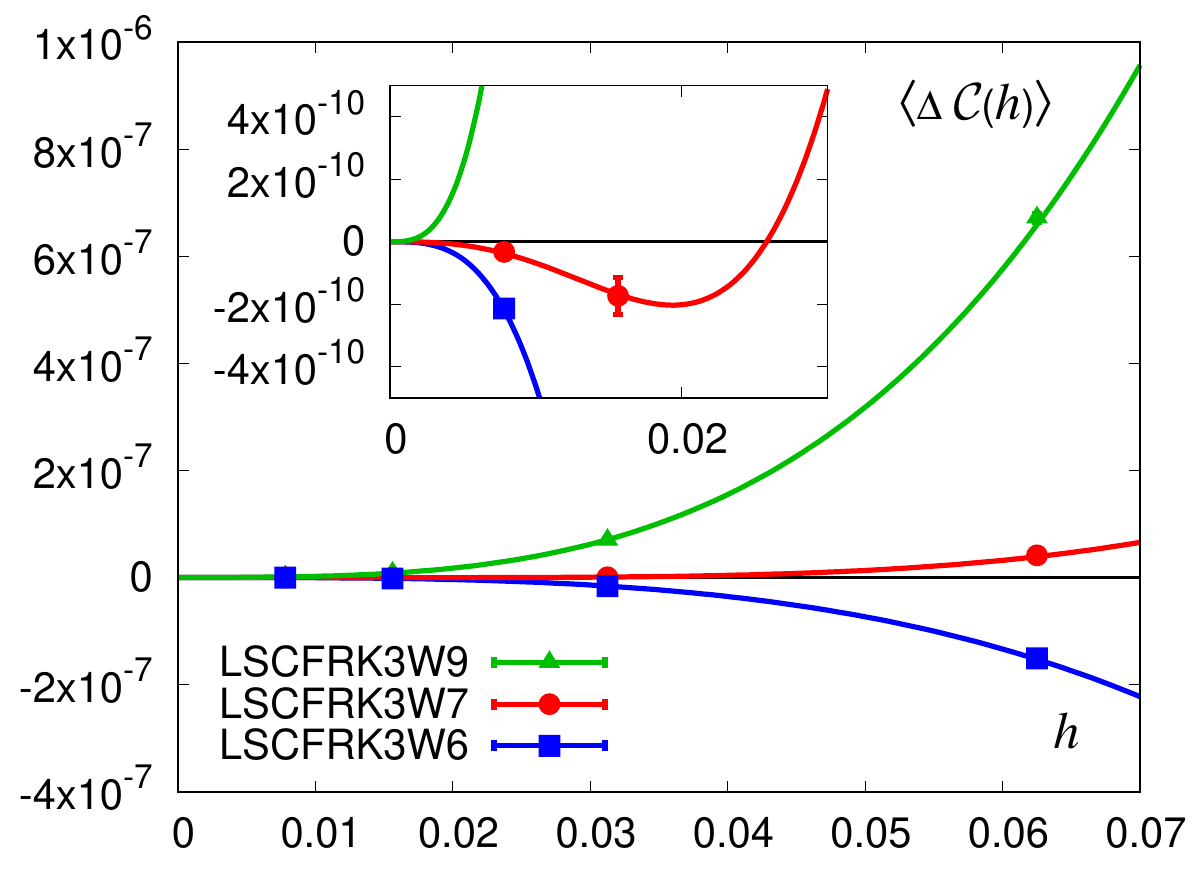}
	\caption{
		The dependence of the global integration error in the clover expression
		$\langle\Delta{\cal C}(h)\rangle$ on the step size $h$ for
		three LSCFRK3 methods. Lines are fits to the data, Eq.~(\ref{eq_DC_fit}).
		For the LSCFRK3W7 method the error changes sign, as shown in the inset.
		\label{fig_clover_3schemes}
	}
\end{figure}

In Fig.~\ref{fig_clover_3schemes} we directly compare
$\langle\Delta{\cal C}(h)\rangle$ for LSCFRK3W6, LSCFRK3W7 and 
LSCFRK3W9\footnote{This scheme has irrational coefficients and it has the
minimal theoretical bound on the global error using the definition
of~\cite{Ralston1962}. In other words, these are ``Ralston coefficients''
but with taking into account the additional constraint of the low-storage
method.
More details can be found
in~\cite{WILLIAMSON198048,Bazavov2020}.}
for the $32^3\times96$ ensemble. The LSCFRK3W7 scheme has the smallest error
$\langle\Delta{\cal C}(h)\rangle$,
which due to the competition between the $h^3$ and $h^4$ terms crosses zero,
as shown in the inset of Fig.~\ref{fig_clover_3schemes}.

To summarize, we expect that the LSCFRK3W7 integrator is close to optimal for
integrating the gradient flow: it has close to the lowest global error for the
gauge fields, Fig.~\ref{fig_norm3}, and its leading-order error coefficient
$D_3^C$ is close to 0 for the action related observables, giving much smaller
global error, Fig.~\ref{fig_clover_3schemes}. Moreover, we observe that these
properties of LSCFRK3W7 are stable against fluctuations within the gauge ensemble,
different ensembles, different types of flow and different discretizations of the
observable.

\begin{figure}[h]
	\centering
	\includegraphics[width=\columnwidth]{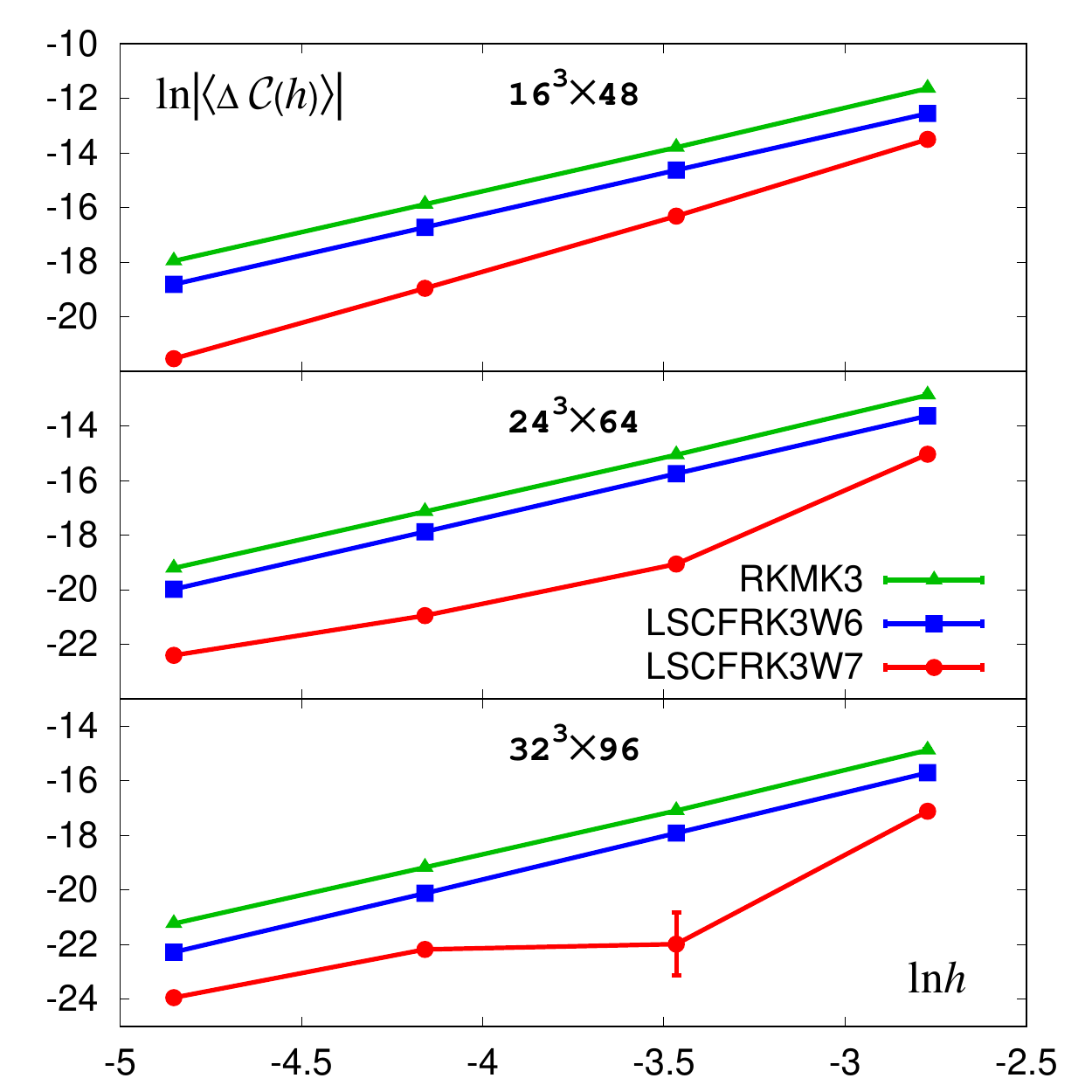}
	\caption{
		Scaling of the global integration error in the clover observable
		$\langle\Delta{\cal C}(h)\rangle$ with the step size $h=1/128$,
		$1/64$, $1/32$ and $1/16$ for the Symanzik flow on the three gauge ensembles.
		The three third-order methods are discussed in the text.
		The lines are drawn to guide the eye.
		\label{fig_scale_rk3}
	}
\end{figure}
In the following we consider only the Symanzik flow and the clover expression
for the observable.
The flow is integrated with LSCFRK3W6 and LSCFRK3W7 on 100 lattices that are
separated by 50 time units for the first two ensembles and 36 time units for the third
in Table~\ref{tab_ensembles}. 
The RKMK8 method with $h=10^{-2}$ is used to get the exact solution.
The error bars are calculated with respect to 20 jackknife bins.
For comparison we also implemented a third-order
algorithm of Munthe-Kaas type with Ralston coefficients,
Algorithm~\ref{alg_RKMK}, that we refer to as
RKMK3. The results for the global integration error for the clover expression
as function of step size $h$
for the three third-order methods are shown in Fig.~\ref{fig_scale_rk3}.

It is customary to plot the logarithm of the absolute value of the error vs
the logarithm of the step size since the slope then is equal to the order of the
method. It also allows one to display the scaling for different integrators
that would be too small on the linear scale. This works well when the error is
dominated by the leading order term. As we observe in Figs.~\ref{fig_clover_zoom},
\ref{fig_clover_3schemes}, the LSCFRK3W7 error crosses zero which translates to
$-\infty$ on the log-log plot. This explains the non-monotonicity of the 
LSCFRK3W7 data in Fig.~\ref{fig_scale_rk3}. The point with the largest error bar
for the $32^3\times96$ lattice is close to the zero crossing and thus the jackknife
propagated error bars in $\ln|\Delta{\cal C}|$ are large there. This is expected
when the leading-order coefficient is small and the leading term is comparable
with the next-to-leading order term for a range of step sizes.
For the $16^3\times48$ lattice the $D_3^C$ coefficient is so close to zero that
the error scales almost as $h^4$ rather than $h^3$. At small enough $h$ all
LSCFRK3 methods should approach the expected $h^3$ behavior.

In Fig.~\ref{fig_scale_rk3} one can observe that the LSCFRK3W7 method has lower
global error than the original integrator of Ref.~\cite{Luscher:2010iy},
LSCFRK3W6. The RKMK3 method has the largest error, despite the fact that its
coefficients are chosen to be the ones that give the lowest
theoretical bound on the global error~\cite{Ralston1962}.

The most expensive part of the calculation is evaluation of the right hand
side $F(Y)$, Eq.~(\ref{eq_dYdt}).
All three methods require three evaluations and thus the
same computational cost. (We neglect the fact that the RKMK3 method also requires
computation of commutators. In a parallel code that operation is local
and thus its overhead is unnoticeable compared with the computation of $F(Y)$,
which requires communication.)
We conclude that LSCFRK3W7 is the most beneficial Lie
group integrator for the gradient flow among the third-order explicit RK methods
that we studied.

\subsection{Properties of two third-order variable step size integrators}
\label{sec_order3_vs}

Variable step size integrators were used for gradient flow in
Refs.~\cite{Fritzsch:2013je,WandeltGuenther2016},
and we also explore methods of this type here for comparison.
In Ref.~\cite{Fritzsch:2013je} a second-order method was embedded into the
LSCFRK3W6 scheme. Methods of this type can be built for all LSCFRK3 schemes so
we consider the most generic case. To connect with the form presented
in~\cite{Fritzsch:2013je} it is convenient to start with the form where
$K_1$, $K_2$ and $K_3$ are separated, Algorithm~\ref{alg_gen}
in~\ref{sec_app_der}. Once the third-order integrator stages are complete and
all $K_i$ are computed, another stage is performed to get a lower order estimate:
\begin{equation}
\label{eq_Y2nd}
\tilde Y_{t+h}=\exp(h(\lambda_3K_3+\lambda_2K_2+\lambda_1K_1))Y_t.
\end{equation}
For $\tilde Y_{t+h}$ to be globally second order (and locally third order) the
coefficients $\lambda_i$ need to satisfy the following conditions:
\begin{eqnarray}
\lambda_1+\lambda_2+\lambda_3&=&1,\\
c_2\lambda_2+c_3\lambda_3&=&\frac{1}{2}.
\end{eqnarray}
For any LSCFRK3 method determined by $c_2(u)$, $c_3(u)$ there is a one-parameter
family of embedded second-order methods. The variable step size scheme is no longer
a low-storage method since $K_i$ need to be stored separately for the
stage~(\ref{eq_Y2nd}) to be applied\footnote{With one exception:
For most methods
there is a single value of $\lambda_3$ where the linear combination
of $K_1$ and $K_2$ can be reused. For instance, for the LSCFRK3W6
method setting $\lambda_3=-1/4$ gives
$\lambda_2K_2+\lambda_1K_1=3\,(8/9K_2-17/36K_1)$, so one can reuse
the same linear combination in the low-order estimate of the solution,
Eq.~(\ref{eq_Y2nd}). And there is an exception to the exception:
A reusable embedded second-order scheme does not exist
for LSCFRK3W7.
See the Mathematica script in \ref{sec_app_math}.}.

For the measure of the local error Ref.~\cite{Fritzsch:2013je} suggested
\begin{equation}
\label{eq_dmax}
d=\max_{x,\mu}\frac{1}{9}\sqrt{{\cal D}(V_{x,\,\mu}(t+h),\tilde V_{x,\,\mu}(t+h))}.
\end{equation}
The full method is summarized in Algorithm~\ref{alg_CFVS} (the local tolerance
$\delta$ is a preset parameter).

\begin{algorithm}[h]
	\caption{4-stage third-order commutator-free variable step size Lie group method}
	\label{alg_CFVS}
	\begin{algorithmic}[1]
		\State Perform the three stages of Algorithm~\ref{alg_gen} for each gauge link
		to get $V_{x,\,\mu}(t+h)$.
		\label{a_s1}
		\State Store $K_i$, $i=1,\dots,3$.
		\State Compute $\tilde V_{x,\,\mu}(t+h)$ using Eq.~(\ref{eq_Y2nd}).
		\State Compute the maximum distance $d$ with Eq.~(\ref{eq_dmax}).
		\State Set $h\to 0.95\sqrt[3]{\frac{\delta}{d}}\,h$.
		\State If $d>\delta$ restart from step \ref{a_s1}.
	\end{algorithmic}
\end{algorithm}
We take $\lambda_3$ as a free parameter (this fixes $\lambda_1$ and $\lambda_2$) and test
two variable step size methods based on $W6$ (as in~\cite{Fritzsch:2013je}) and $W7$
coefficients. These methods are referred to as CFRK3W6VS and CFRK3W7VS in the following
(VS = variable step size). The tests are performed on a \textit{single}
$16^3\times48$ lattice
with the Symanzik flow to illustrate a qualitative point.

\begin{figure}[h]
	\centering
	\includegraphics[width=\columnwidth]{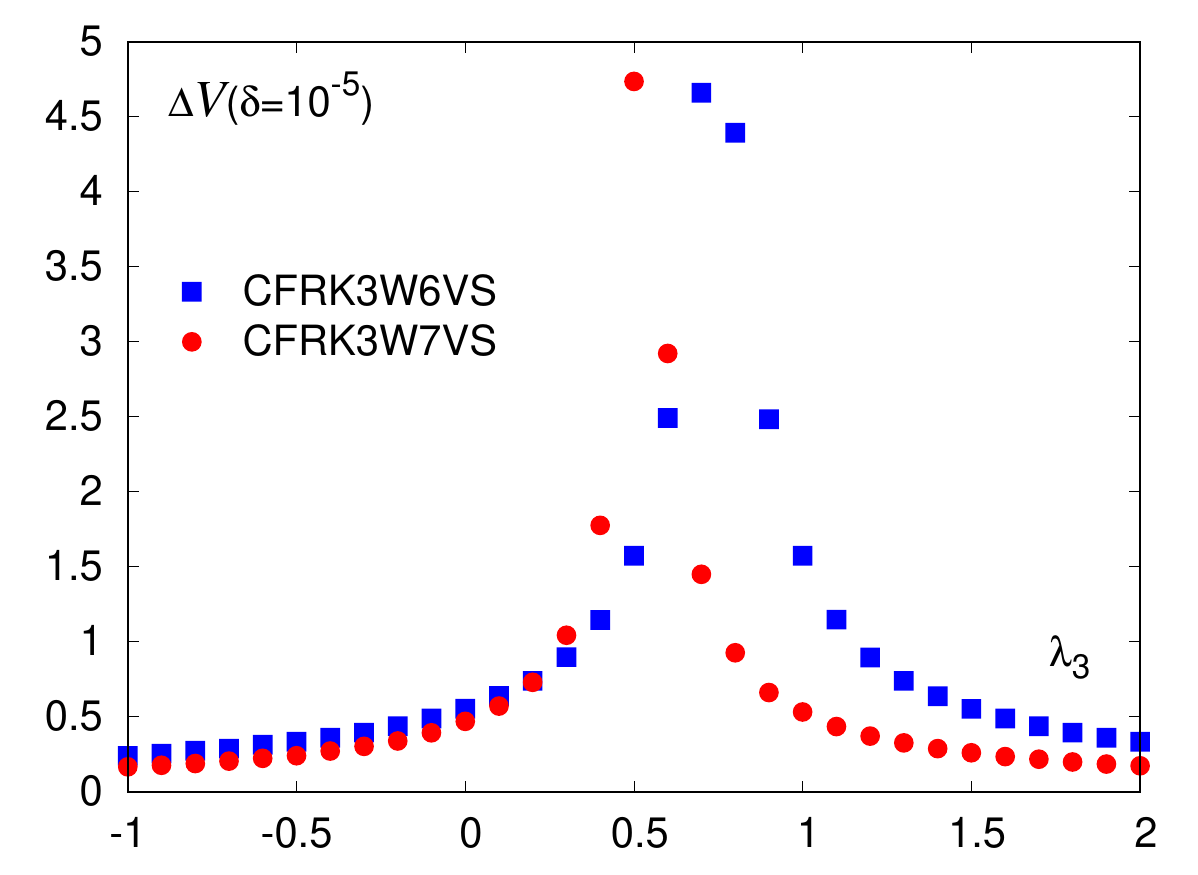}
	\caption{
		The global integration error for gauge fields defined in Eq.~(\ref{eq_dist_norm})
		as function of the parameter $\lambda_3$ that distinguishes variable step
		size methods. The CFRK3W6VS and CFRK3W7VS methods differ in what third-order
		integrator is used. The global error is evaluated at the same local
		tolerance $\delta=10^{-5}$ for all methods.
		The results are from the flow on a single $16^3\times48$ lattice
		and therefore have no errorbars.
		\label{fig_rkvs_all}
	}
\end{figure}

\begin{figure}[h]
	\centering
	\includegraphics[width=\columnwidth]{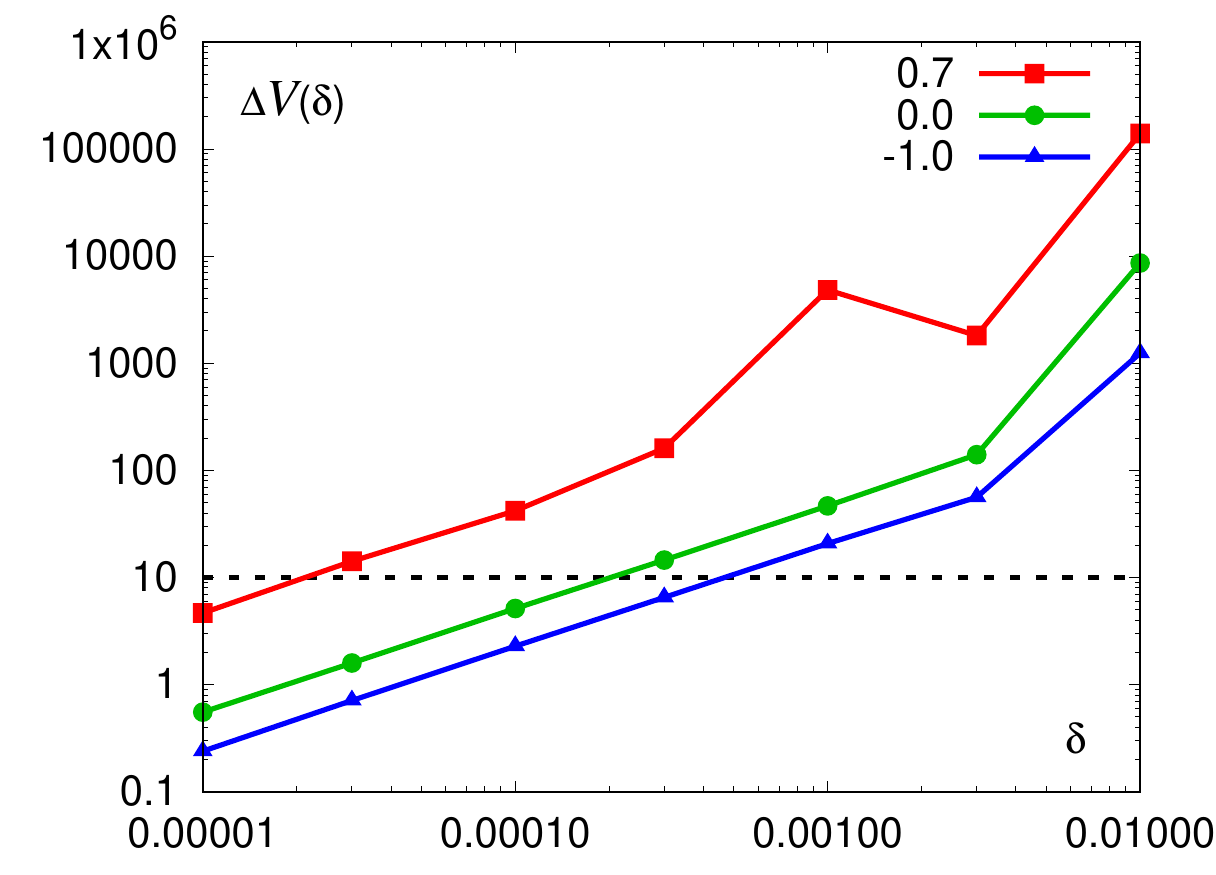}
	\caption{
		The global integration error $\Delta V(\delta)$ as function of the local
		tolerance $\delta$ for the three third-order variable step size methods
		CFRK3W6VS($\lambda_3$) for $\lambda_3=-1$, $0$ and $0.7$.
		The results are from the flow on a single $16^3\times48$ lattice and therefore
		have no errorbars.
		\label{fig_rkvs3}
	}
\end{figure}
For a set of local tolerances $\delta=10^{-5}$, $3\times10^{-5}$,
$10^{-4}$, $3\times10^{-4}$, $10^{-3}$, $3\times10^{-3}$ and $10^{-2}$
CFRK3W6VS and CFRK3W7VS were run for values of $\lambda_3\in[-1,2]$ separated by $0.1$.
The results for the global error in the gauge fields $\Delta V(\delta)$
defined in~(\ref{eq_dist_norm})
are shown in Fig.~\ref{fig_rkvs_all} for a fixed value of tolerance $\delta=10^{-5}$
as function of the $\lambda_3$ parameter. There seems to be some room for tuning $\lambda_3$
since there is a significant difference in what global error is achieved by different
methods. In Fig.~\ref{fig_rkvs3} the dependence of the global error
$\Delta V(\delta)$ on the local tolerance $\delta$ is shown for the three
CFRK3W6VS methods with $\lambda_3=-1$, $0$ and $0.7$. Note that $\lambda_3=0$
corresponds to the variable step size method of Ref.~\cite{Fritzsch:2013je}.
$\lambda_3=0.7$ gives the highest error among the CFRK3W6VS schemes shown in
Fig.~\ref{fig_rkvs_all}.
It may seem
that $\lambda_3=-1$ is the best method.

\begin{figure}[h]
	\centering
	\includegraphics[width=\columnwidth]{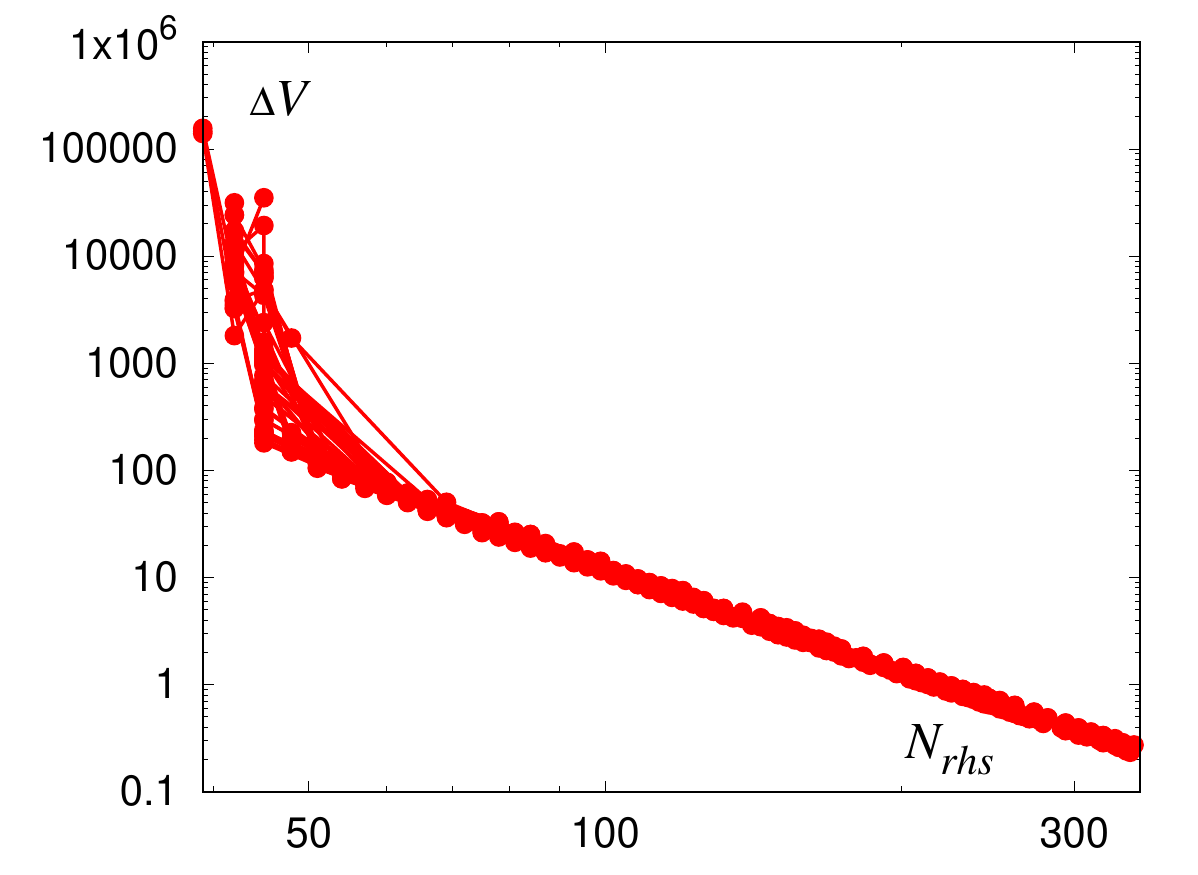}
	\caption{
		The global error $\Delta V$ as function of the number of right hand side
		evaluations of Eq.~(\ref{eq_flow}) for all 62 variable step size schemes
		shown in Fig.~\ref{fig_rkvs_all}.
		The results are from the flow on a single $16^3\times48$ lattice
		and therefore have no errorbars.
		\label{fig_rkvs_rhs}
	}
\end{figure}
However, the relevant question to address is what number of steps (\textit{i.e.}
computational effort) each method needs to reach a certain global error.
For this purpose all 62 schemes shown in Fig.~\ref{fig_rkvs_all} are plotted against
the number of right hand side evaluations (which is the number of steps times three
stages per step) $N_{rhs}$ in Fig.~\ref{fig_rkvs_rhs}.
All schemes collapse on the same line apart from the region of small
number of steps at the beginning (large local tolerance) where the subleading corrections
to scaling are still large. There is no room for tuning -- all methods would reach a given
global error with the same computational effort independently of what coefficients of
the embedded low-order scheme $\lambda_i$ are used. What is different for different
schemes is the value of the local tolerance at which that global error is achieved.
For instance, the CFRK3W6VS with $\lambda_3=0.7$ achieves the global error
$\Delta V=10$ at the local tolerance $\delta=2.2\times10^{-5}$, with
$\lambda_3=0$ at $\delta=2\times10^{-4}$ and with $\lambda_3=-1$ at
$\delta=4.5\times10^{-4}$, Fig.~\ref{fig_rkvs3}.

This brings us to an important point. It may be tempting, as happens in some literature,
to interpret the local tolerance $\delta$ as a universal parameter that can be set
once and after that the integrator self-tunes and takes care of the global error.
Contrary to that, we observe that there is no apriori way to know what global integration
error is achieved for a specific value of $\delta$ on a given lattice ensemble.
Therefore, $\delta$ should be treated in the same way as the step size $h$ in fixed
step size methods. For a calculation performed at a single value of $h$ there is no way
to estimate what global error was achieved, apart from knowing that it is proportional
to $h^p$, where $p$ is the order of the method. One needs to either calculate the
exact (or high-precision) solution and compare with it, or perform the calculation at
several step sizes, study the dependence of the error in the observable of interest
on the step size and decide what step size is appropriate for a given lattice ensemble.
Similarly, one needs to study the scaling of the error with respect to $\delta$ and
pick $\delta$ based on that. It will certainly be different for each problem, since
the numerical value of the global integration error is influenced by many factors
such as the integration method, what gauge ensembles are used, type of the gradient flow,
for how long the flow is run, etc.

Given that all embedded methods are equivalent with respect to the computational cost
for a given higher-order scheme and we do not observe a significant difference between
CFRK3W6VS and CFRK3W7VS\footnote{This happens also because W6 and W7 schemes are close,
\textit{e.g.} Fig.~\ref{fig_w_rat}. For a method further away on the Williamson curve
the embedded schemes would still be equivalent themselves, but the computational
cost of that integrator would be higher than for CFRK3W6VS and CFRK3W7VS.},
we only perform calculations with the original method of Ref.~\cite{Fritzsch:2013je} which is
CFRK3W6VS($\lambda_3=0$).

For comparison we also implemented the Bogacki-Shampine variable step size
integrator~\cite{BOGACKI1989321} that was used for gradient flow in 
Ref.~\cite{WandeltGuenther2016}. The Lie group integrator
of this type is built as an extension of the
RKMK3 method (\textit{i.e.} involves commutators). It requires four right hand side
evaluations; however, the last evaluation at the current step is the first on the next,
so it can be stored and reused (so called First Same As Last -- FSAL property).
Therefore in practice it requires only three right hand side evaluations, which is the
same computational effort as for all the other third-order methods explored here.
Since we do not find this method to be beneficial, we do not list the full algorithm.
The details can be found in~\cite{BOGACKI1989321,WandeltGuenther2016}.
This method is referred to as RKMK3BS.

\begin{figure}[h]
	\centering
	\includegraphics[width=\columnwidth]{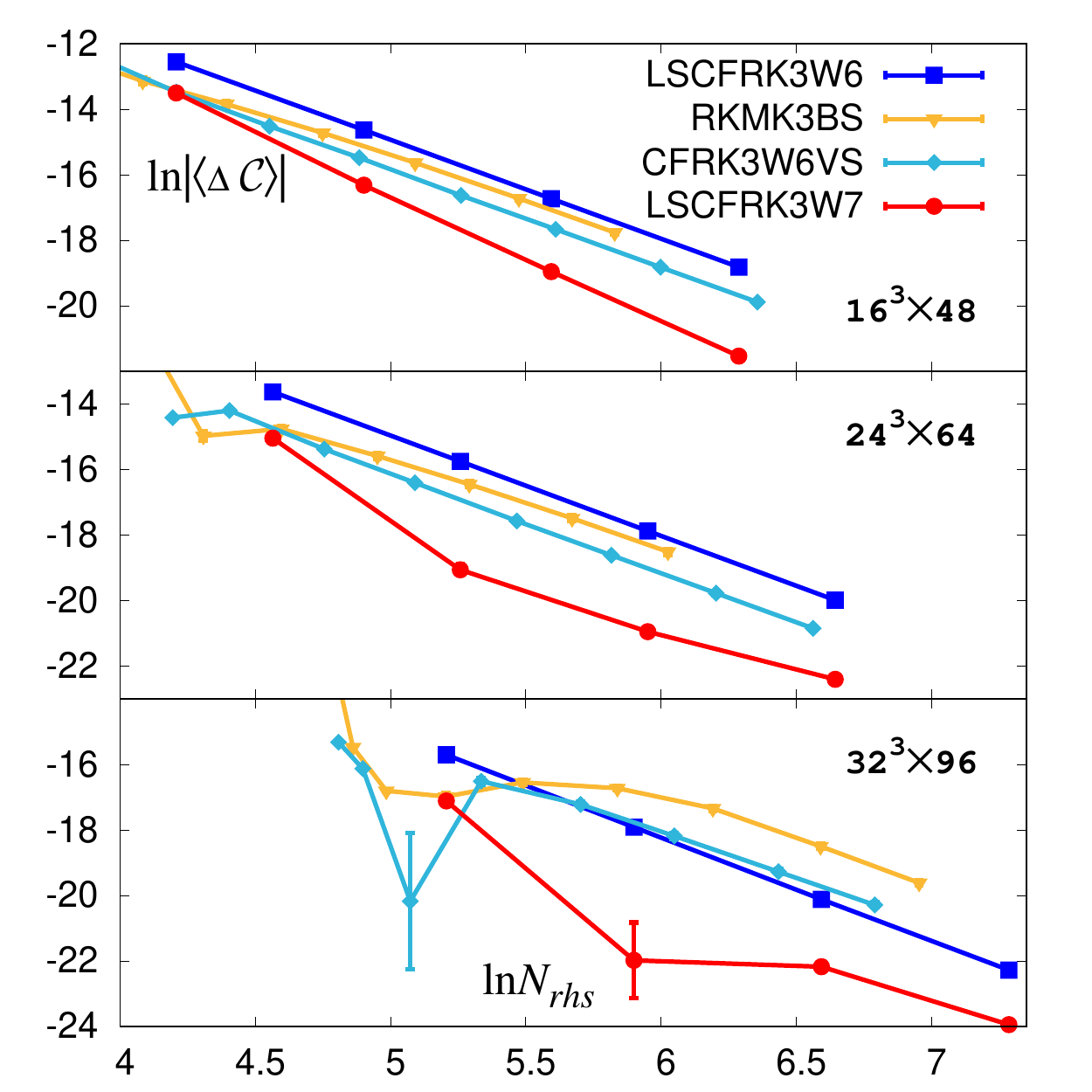}
	\caption{
		Scaling of the global integration error in the clover observable
		$\langle\Delta{\cal C}\rangle$ with the number of right hand side evaluations
		$N_{rhs}$ for the Symanzik flow on the three gauge ensembles.
		The variable step size methods RKMK3BS and CFRK3W6VS were run
		with a set of local tolerances $\delta=10^{-6}$,
		$3\times10^{-6}$, $10^{-5}$, $3\times10^{-5}$, $10^{-4}$, $3\times10^{-4}$, $10^{-3}$, $3\times10^{-3}$. The data for the fixed step size
		methods LSCFRK3W6 and LSCFRK3W7 is the same as in
		Fig.~\ref{fig_scale_rk3}.
		The lines are drawn to guide the eye.
		\label{fig_scale_rk3vs}
	}
\end{figure}
As in Sec.~\ref{sec_order3} the Symanzik flow is measured on 100 lattices from each
ensemble of Table~\ref{tab_ensembles}. The error bars are estimated from
20 jackknife bins.
In Fig.~\ref{fig_scale_rk3vs} the logarithm of the global error
$\langle\Delta{\cal C}(h)\rangle$
is shown vs the logarithm of the number of the right hand side evaluations
$N_{rhs}$ for the two fixed step size integrators LSCFRK3W6 and
LSCFRK3W7 that are discussed in Sec.~\ref{sec_order3} and the two variable
step size methods CFRK3W6VS($\lambda_3=0$) and RKMK3BS. The origin of the non-monotonic
behavior for CFRK3W6VS is similar to the one for LSCFRK3W7 -- its global error
crosses zero at some value of $\delta$. Interestingly, while both variable step size
methods are more beneficial than LSCFRK3W6 for the $16^3\times48$ and $24^3\times64$
ensembles, for the $32^3\times96$ ensemble in the regime where all integrators
approach the expected cubic scaling, the variable step size schemes require comparable
or larger computational effort than LSCFRK3W6. The fixed step size LSCFRK3W7
method requires the least computational effort for all three ensembles except
the region of large local tolerance (small number of right hand side
evaluations). We conclude
that for the scale setting applications there may be no benefit from using variable
step size integrators, at least for the gauge ensembles that we used for this study.

Our findings are in contrast with the studies of Ref.~\cite{Fritzsch:2013je} where 
large computational savings were reported for CFRK3W6VS($\lambda_3=0$). However,
it appears that there the gradient flow was used in a very different regime.
Apart from less important differences in the gauge couplings,
boundary conditions and lattice volumes, the flow was run significantly longer
to achieve a much larger smoothing radius than is needed for $w_0$-scale
setting. Translated for the gauge ensembles used in this study the flow 
would be run in the ranges $T_{max}=2.9-8$ for the $a=0.15$~fm,
$T_{max}=6.5-18$ for the $a=0.12$~fm and
$T_{max}=11.5-32$ for the $a=0.09$~fm ensembles
(compare with $T_{max}$ in Table~\ref{tab_ensembles}).

\subsection{Two low-storage fourth-order Lie group integrators}
\label{sec_order4}

Reference~\cite{Bazavov2020} opened a possibility of building low-storage Lie group
integrators from classical $2N$-storage RK methods. There are a number of fourth-order
$2N$-storage methods in the literature~\cite{CK1994,BERNARDINI20094182,BERLAND20061459,TOULORGE20122067,STANESCU1998674,ALLAMPALLI20093837,NIEGEMANN2012364} that were constructed for different applications.
Most of them include large number of stages and may therefore be computationally 
inefficient for integration of the gradient flow in lattice gauge theory.

We study two fourth-order methods that have five
(the minimal possible number)~\cite{CK1994} and six~\cite{BERLAND20061459} stages
which we refer to as LSCFRK4CK and LSCFRK4BBB\footnote{We use
nomenclature consistent with other methods in this paper. In the original
work~\cite{BERLAND20061459} the classical Runge-Kutta method is called 
RK46-NL.}, respectively.
Their coefficients were found by solving a system of nonlinear equations
resulting from eight order conditions for a classical RK method and additional
problem-dependent constraints (\textit{e.g.} increased regions of stability). These
coefficients in the $2N$-storage format are listed in \ref{sec_app_rw}.

As in Sec.~\ref{sec_order3}, the tests are performed with the Symanzik flow,
clover observable on 100 lattices from the ensembles listed in Table~\ref{tab_ensembles}.
The step sizes $h=1/16$, $1/32$, $1/64$ and $1/128$ are used and the exact solution
is obtained with RKMK8 at $h=10^{-2}$. For comparison we also implemented
a fourth-order method of Munthe-Kaas type, RKMK4.
Such a method was employed
in~\cite{Ce:2015qha}. Our implementation is not exactly the same as there
and provides a slightly smaller global error, but
our findings are qualitatively similar.

\begin{figure}[h]
	\centering
	\includegraphics[width=\columnwidth]{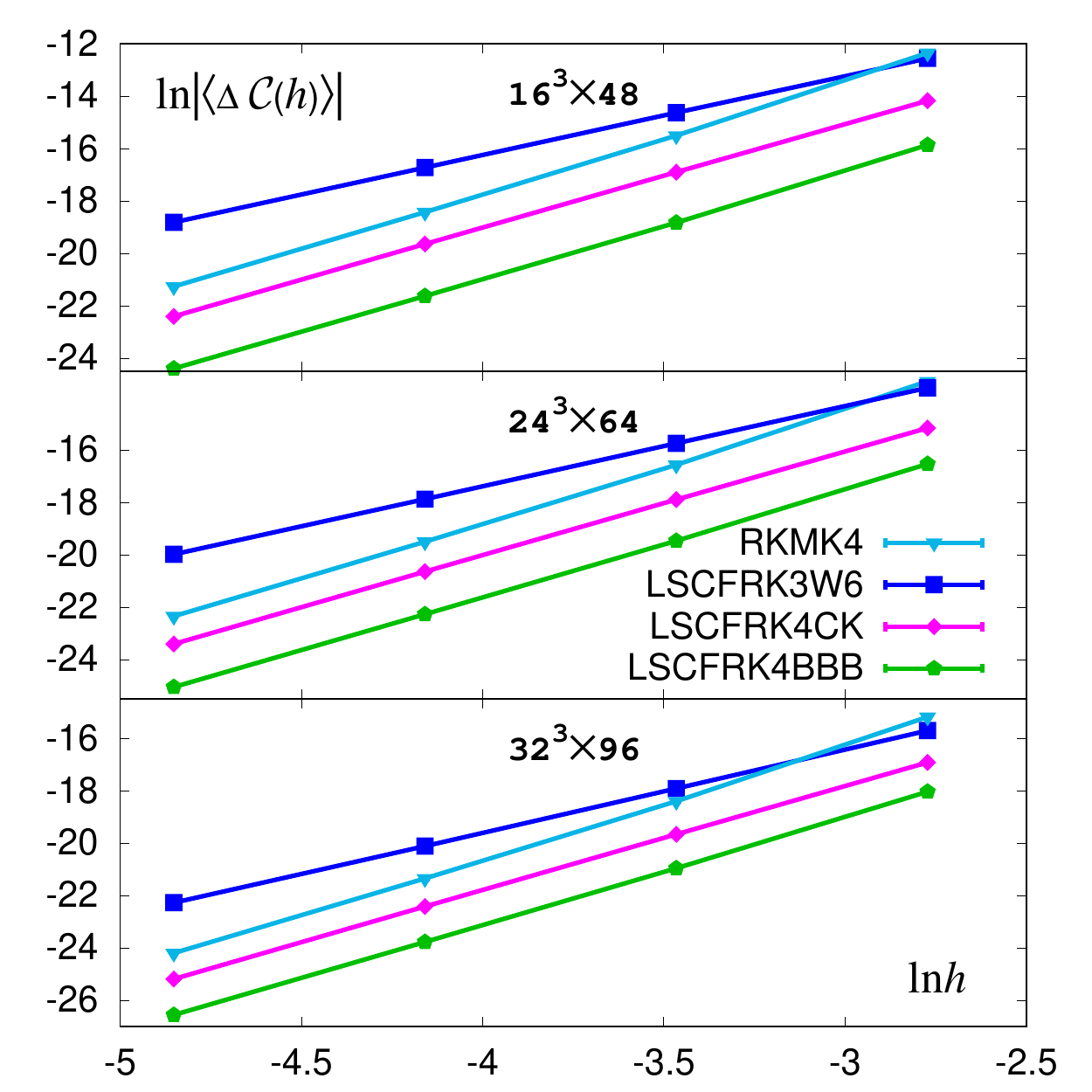}
	\caption{
		Scaling of the global integration error in the clover observable
		$\langle\Delta{\cal C}(h)\rangle$ with the step size $h=1/128$,
		$1/64$, $1/32$ and $1/16$ for the Symanzik flow on the three gauge ensembles.
		The three fourth-order methods are discussed in the text.
		The lines are drawn to guide the eye.
		\label{fig_scale_rk4}
	}
\end{figure}
The scaling of the error in the clover observable $\ln|\langle\Delta{\cal C}(h)\rangle|$
as function of $\ln h$ is shown in Fig.~\ref{fig_scale_rk4} for the LSCFRK3W6,
LSCFRK4CK, LSCFRK4BBB and RKMK4 methods. RKMK4 has the largest error among the
fourth-order methods while the lowest global integration error is achieved with
LSCFRK4BBB.

The $2N$-storage RK methods with $s$ stages have $2s-1$ parameters
($A_1=0$ for explicit methods). Thus, the five-stage
LSCFRK4CK integrator also belongs to
a one-parameter family since there are 8 classical order conditions at fourth
order. There may also be some possibility for tuning that integrator similarly to
our discussion in Sec.~\ref{sec_order3}. However, due to the complexity of
the order conditions, no analytic solutions such as Eq.~(\ref{eq_Wc2c3}) are
available. This makes tuning of that integrator a complicated task. We note
that there are three more coefficient schemes reported in Ref.~\cite{CK1994}, but
we found them less efficient than the main scheme that Ref.~\cite{CK1994}
recommended and which was implemented in this study.

\subsection{Final comparison}

\begin{figure}
	\centering
	\includegraphics[width=\columnwidth]{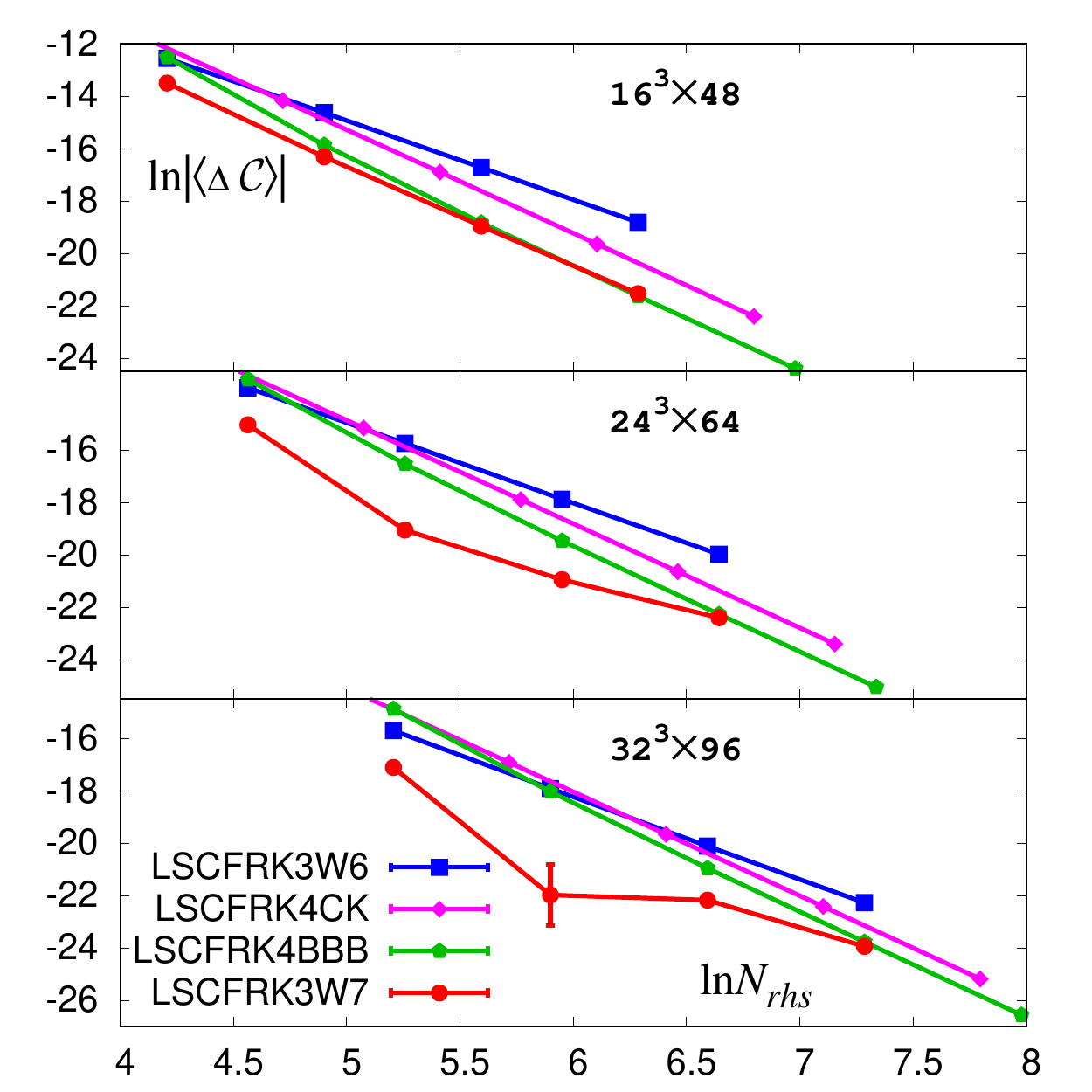}
	\caption{
		Scaling of the global integration error in the clover observable
		$\langle\Delta{\cal C}\rangle$ with the number of right hand side evaluations
		$N_{rhs}$ for the Symanzik flow on the three gauge ensembles.
		All integrators were run at step sizes $h=1/128$,
		$1/64$, $1/32$ and $1/16$ except LSCFRK4CK and
		LSCFRK4BBB where $h=1/8$ was also included.
		The lines are drawn to guide the eye.
		\label{fig_scale_rk34}
	}
\end{figure}

The integration schemes that we explored differ in the number of stages.
To compare their computational efficiency,
Fig.~\ref{fig_scale_rk34} shows
the dependence of the global error
in the clover observable vs the number of right hand side evaluations
for the third-order method of Ref.~\cite{Luscher:2010iy}
LSCFRK3W6, the third-order scheme LSCFRK3W7 that we discussed in Sec.~\ref{sec_order3},
and the two fourth-order methods LSCFRK4CK and LSCFRK4BBB.
Compared with LSCFRK3W6 we find that the LSCFRK3W7 scheme produces lower global error
at all step sizes explored and is thus more beneficial computationally.
The fourth-order LSCFRK4BBB scheme becomes more beneficial than LSCFRK3W6
at step size of about $h=1/16$ for the $16^3\times48$ and $24^3\times64$ ensembles
and about $h=1/32$ for the $32^3\times96$ ensemble (the step size here is for
the LSCFRK3W6 integrator, the one for LSCFRK4BBB is about twice as large at the
crossing point). Compared with LSCFRK3W7, LSCFRK4BBB becomes more efficient at $h=1/128$
($h$ for LSCFRK3W7). For the $32^3\times96$ ensemble the five-stage LSCFRK4CK
method becomes comparable with LSCFRK4BBB.

\section{Conclusion}
\label{sec_concl}
Based on the connection~\cite{Bazavov2020}
between the $2N$-storage classical Runge-Kutta
methods~\cite{WILLIAMSON198048}
and commutator-free integrators~\cite{CELLEDONI2003341} we explored several
possible improvements in the efficiency of integrating the gradient flow in
lattice gauge theory.

Among the low-storage three-stage third-order schemes that are parametrized by
the Williamson curve the LSCFRK3W7 is the most promising. Its global error in the
norm of the gauge field is close to the minimum, Fig.~\ref{fig_norm3}.
For the action observables this method is close to the point where the
leading order error coefficient is close to zero. Like the originally
proposed LSCFRK3W6 method of L\"{u}scher~\cite{Luscher:2010iy}, LSCFRK3W7
has ration\-al coefficients that are given in the $2N$-storage form in
\ref{sec_app_rw}.
The performance of the LSCFRK3W7 method is universal across ensembles
with different lattice spacing,
types of flow and types of observable that we explored.
For a specific gradient flow application the reader can 
always revisit the tuning of the
third-order scheme similar to our study in Sec.~\ref{sec_order3}
by \textit{e.g.} running a small-scale test on the set of
rational values of $c_2$ and $c_3$ coefficients given in~\ref{sec_app_rw}
that reasonably cover the Williamson curve. For the reader's convenience
we also provide a listing of the Mathematica script
that calculates the LSCFRK3 method coefficients in various formats from given
$c_2$ and $c_3$ in \ref{sec_app_math}.

Our studies of the third-order variable step size methods in Sec.~\ref{sec_order3_vs}
indicate that one needs to exercise caution in interpreting the local tolerance
$\delta$ parameter. Its relation to the global integration error is not
\textit{a priori}
known, in the same way as it happens with the step size $h$ for fixed step size methods.
Thus, in both situations one needs to study the scaling of the global
error with the control parameter, step size $h$ or local tolerance $\delta$,
and tune it based on that for each specific case. For $w_0$-scale setting applications
we find that it is still computationally more efficient to use the fixed step size
LSCFRK3W7 method rather than the third-order variable step size schemes.
For applications where the third-order variable step size methods
may be beneficial we also include the coefficient scheme that allows one to
reuse the second stage of the third-order integrator in the embedded
second-order integrator in the Mathematica script in \ref{sec_app_math}.

There are two low-storage fourth-order methods studied in Sec.~\ref{sec_order4}
that may be well-suited for gradient flow applications. We find that the
LSCFRK4BBB method is the most computationally efficient one, although the
five-stage LSCFRK4CK method becomes comparable at finer ensembles. For the gauge
ensembles that we studied we conclude that if one needs to run the LSCFRK3W6
integrator at time steps lower that $1/32$, it is more beneficial to switch
to LSCFRK4BBB. 
A fourth-order RKMK4 method was used
in Ref.~\cite{Ce:2015qha} to
provide a conservative estimate of the integration error for observables related
to the topology of gauge fields. We believe that LSCFRK4BBB provides a better
alternative to RKMK4, see Fig.~\ref{fig_scale_rk4}.
For the three gauge ensembles used here the LSCFRK4BBB
integrator is stable at the largest step size we tried, $h=1/8$. The average
integration error at this step size for the clover observable is
$\langle \Delta{\cal C}(h=1/8)\rangle=-3.7\times10^{-6}$ for the $a=0.15$~fm,
$\langle \Delta{\cal C}(h=1/8)\rangle=-1.7\times10^{-6}$ for the $a=0.12$~fm and
$\langle \Delta{\cal C}(h=1/8)\rangle=-3.5\times10^{-7}$ for the $a=0.09$~fm ensemble.

Based on our findings we recommend LSCFRK3W7, LSCFRK4BBB and LSCFRK4CK for
integrating the gradient flow.
Exact speedups depend on the accuracy required for the flow observables.
For the ensembles we used LSCFRK3W7 is about twice as efficient
as LSCFRK3W6 for the $w_0$-scale setting.
Since these methods are all written in the
$2N$-storage format, it is very easy to implement different integrators
in the existing code. For instance, in the MILC code~\cite{MILCgit}
the main loop over the stages in Algorithm~\ref{alg_CF_2N_RK} is the
following\footnote{The minus sign in front of the second argument is to match
the historic convention in the MILC code on how the right hand side of
Eq.~(\ref{eq_flow}) is computed. Also, at the time of writing the code
for all integrators used in this paper is located in the \texttt{feature/wilson\_flow\_2} branch.}:
\begin{verbatim}
  for( i=0; i<N_stages; i++ ) {
    integrate_RK_2N_one_stage( A_2N[i],
                  -B_2N[i]*stepsize );
  }
\end{verbatim}
The only change required in the code to implement a different LSCFRK integrator
is to change the following compile-time parameters:
\texttt{N\_stages} -- the number of stages of the method, and
\texttt{A\_2N}, \texttt{B\_2N} -- the values of the two arrays of size \texttt{N\_stages}
that store the $A_i$, $B_i$ coefficients in the $2N$-storage format.
The code in the \texttt{integrate\_RK\_2N\_one\_stage}
function performs one
stage of Algorithm~\ref{alg_CF_2N_RK} and is typically already present
in the lattice codes that implemented Algorithm~\ref{alg_W6}.

To summarize, we presented several Lie group methods that may provide better
computational efficiency for integrating the gradient flow in lattice gauge theory.
Given their low-storage properties, they are easy to implement in the existing codes.
For the reader interested in exploring the properties and implementation of the low-storage
Lie group integrators in a simpler setting we point out that a simple Matlab script
for integrating the equation of motion of a rotating free rigid body is included in the
Appendix of Ref.~\cite{Bazavov2020}.

\bigskip
\textbf{Acknowledgements.} 
We thank the MILC collaboration for sharing the gauge configurations,
Oliver Witzel for an independent test of the flow observables,
Oswald Knoth for checking the order conditions for the low-storage Lie group
integrators used here and
Claude Bernard, Steven Gottlieb and Johannes Weber
for careful reading and comments on the manuscript.
This work was in part supported
by the U.S. National Science Foundation under award
PHY-1812332.

\appendix

\section{Some order conditions for classical Runge-Kutta methods}
\label{sec_app_oc}
The coefficients of a third-order explicit classical Runge-Kutta method satisfy
the following order conditions:
\begin{eqnarray}
\sum_i b_i &=& 1,\\
\sum_i b_ic_i &=& \frac{1}{2},\\
\sum_i b_ic_i^2 &=& \frac{1}{3},\\
\sum_{i,j} b_i a_{ij} c_j &=& \frac{1}{6}.
\end{eqnarray}
A 3-stage method has six independent parameters.
Picking $c_2$ and $c_3$ as free parameters one gets
the most generic branch of solutions
$c_2\neq 0 \neq c_3\neq c_2\neq 2/3$~\cite{ButcherBook}:
\begin{eqnarray}
b_2 &=& \frac{3c_3-2}{6c_2(c_3-c_2)},\label{eq_b2}\\
b_3 &=& \frac{2-3c_2}{6c_3(c_3-c_2)},\\
a_{32} &=& \frac{c_3(c_3-c_2)}{c_2(2-3c_2)},\\
b_1 &=& 1-b_2-b_3,\\
a_{21} &=& c_2,\\
a_{31} &=& c_3-a_{32}.\label{eq_a31}
\end{eqnarray}

\section{Coefficients for several $2N$-storage third- and fourth-order classical RK methods}
\label{sec_app_rw}
For future reference and possible integrator tuning we list several rational
coefficient schemes in Table~\ref{tab_rat_w}.

\begin{table}
	\centering
	\caption{
		Three-stage third-order $2N$-storage schemes with rational coefficients.
		The original numbering of Ref.~\cite{WILLIAMSON198048} is listed in the
		first column. The additional schemes that we found are marked with the
		asterisk. There are many more than we show here,
		but as the numerator and denominator
		grow larger they are not convenient to use.
		\label{tab_rat_w}
	}
\vspace{8pt}
\begin{tabular}{rrr}
Numbering~\cite{WILLIAMSON198048} & $c_2$ & $c_3$  \\
\hline
4  & $2/3$ & $0$ \\
*  & $7/12$ & $2/15$ \\
*  & $823/1887$ & $153/592$ \\
5  & $1/4$ & $5/12$ \\
*  & $1418/6783$ & $9894/19285$ \\
*  & $1418/6783$ & $1064/1887$ \\
6  & $1/4$ & $2/3$ \\
7  & $1/3$ & $3/4$ \\
*  & $823/1887$ & $5365/6783$ \\
*  & $9391/19285$ & $5365/6783$ \\
*  & $7/12$ & $3/4$ \\
*  & $439/592$ & $1064/1887$ \\
*  & $13/15$ & $5/12$ \\
12  & $2/3$ & $2/3$ \\
14  & $1$ & $1/3$ \\
\end{tabular}
\end{table}

The coefficients of the RK3W6 and RK3W7 schemes in the $2N$-storage format
are listed in Table~\ref{tab_2N} and of the RK4CK and RK4BBB in
Table~\ref{tab_2N_4}. Note that the RK4CK coefficients were found
numerically and then close rational expressions were found that provide
26 digits of accuracy. For the RK4BBB method the coefficients are given
with 12 digits of accuracy in~\cite{BERLAND20061459} therefore one cannot
expect the error to be below $10^{-12}$. This bound
is however well below the typical precision needed in gradient flow
applications.

\begin{table}
	\centering
	\caption{
		The coefficients of the two $2N$-storage third-order schemes 
		of~\cite{WILLIAMSON198048}
		that are the basis for the LSCFRK3W6 and LSCFRK3W7 methods
		discussed in Sec.~\ref{sec_order3}.
		\label{tab_2N}
	}
	\vspace{8pt}
	\begin{tabular}{rrr}
		Coefficient & RK3W6 & RK3W7  \\
		\hline
		$A_1$  & $0$ & $0$ \\
		$A_2$  & $-17/32$ & $-5/9$ \\
		$A_3$  & $-32/27$ & $-153/128$ \\
		$B_1$  & $1/4$ & $1/3$ \\
		$B_2$  & $8/9$ & $15/16$ \\
		$B_3$  & $3/4$ & $8/15$ \\
	\end{tabular}
\end{table}

\begin{table}
	\centering
	\caption{
		The coefficients of the two $2N$-storage fourth-order schemes 
		of~\cite{CK1994} and \cite{BERLAND20061459}
		that are the basis for the LSCFRK4CK and LSCFRK4BBB methods
		discussed in Sec.~\ref{sec_order4}.
		\label{tab_2N_4}
	}
	\vspace{8pt}
	\begin{tabular}{rr}
		Coefficient & RK4CK \\
		\hline
		$A_1$  & $0$ \\
		$A_2$  & $-567301805773/1357537059087$  \\
		$A_3$  & $-2404267990393/2016746695238$ \\
		$A_4$  & $-3550918686646/2091501179385$ \\
		$A_5$  & $-1275806237668/842570457699$  \\
		$B_1$  & $1432997174477/9575080441755$  \\
		$B_2$  & $5161836677717/13612068292357$ \\
		$B_3$  & $1720146321549/2090206949498$  \\
		$B_4$  & $3134564353537/4481467310338$  \\
		$B_5$  & $2277821191437/14882151754819$ \\
		\hline
		Coefficient & RK4BBB \\
		\hline
		$A_1$  & $0$ \\
		$A_2$  & $-0.737101392796$ \\
		$A_3$  & $-1.634740794341$ \\
		$A_4$  & $-0.744739003780$ \\
		$A_5$  & $-1.469897351522$ \\
		$A_6$  & $-2.813971388035$ \\
		$B_1$  & $0.032918605146$ \\
		$B_2$  & $0.823256998200$ \\
		$B_3$  & $0.381530948900$ \\
		$B_4$  & $0.200092213184$ \\
		$B_5$  & $1.718581042715$ \\
		$B_6$  & $0.27$ \\
	\end{tabular}
\end{table}

\section{Derivation of the Lie group integrator of Ref.~\cite{Luscher:2010iy}}
\label{sec_app_der}
It is illustrative to discuss the properties of the integrator provided by
L\"{u}scher in Ref.~\cite{Luscher:2010iy}. As this coefficient scheme was not present
in the literature on Lie group methods at the time, we believe that the method
was derived independently. Let us start with a generic algorithm that
reuses the function value from the previous stage, Algorithm~\ref{alg_gen},
and contains six yet undetermined parameters.

\begin{algorithm}[h]
	\caption{3-stage generic Lie group method with reuse of function values}
	\label{alg_gen}
	\begin{algorithmic}[1]
		\State $Y_1=Y_t$
		\State $K_1=F(Y_1)$
		\State $Y_2=\exp\left(h\alpha_{21}K_1\right)Y_1$
		\State $K_2=F(Y_2)$
		\State $Y_3=\exp\left(h\left(\alpha_{32}K_2+\alpha_{31}K_1\right)\right)Y_2$
		\State $K_3=F(Y_3)$
		\State $Y_{t+h} = \exp\left(h\left(\beta_{3}K_3+\beta_{2}K_2+\beta_{1}K_1\right)\right)Y_3$
		\Statex \textbf{or}
        \setcounter{ALG@line}{6} 
        {\let\fmtlinenumber\mathspecalg
		\State $Y_{t+h} = \exp\left(h\left(\beta_{3}K_3+c(\alpha_{32}K_2+\alpha_{31}K_1)\right)\right)Y_3$
		\Comment{See text.}
	}
	\end{algorithmic}
\end{algorithm}

If $Y_3$ and $Y_2$ are substituted
in terms of $Y_1$, one can recognize that this scheme belongs to the class of
commutator-free Lie group integrators explored in Ref.~\cite{CELLEDONI2003341}.
However, it does not belong to the classes of solutions
found there.

We would like to find a solution of the order conditions
that allows for a third-order global accuracy method.
For a classical RK method there are four order conditions,
thus, we need at least four parameters to satisfy them.
For a Lie group integrator of this type we need at least five parameters.
This can be seen from the following argument.
If we take, for instance, a third-order Crouch-Grossman method~\cite{Crouch1993}
(which is a commutator-free Lie group integrator with a more restricted structure
than the methods of~\cite{CELLEDONI2003341}),
there is a non-classical order condition arising from non-commutativity,
so there are five in total.
If we follow the RKMK route~\cite{MuntheKaas1998,MUNTHEKAAS1999115},
one needs, at least, one commutator for a third-order method.
Thus, if we would like to avoid commutators,
we need at least one more coefficient to tune
to cancel the effect of the commutator.
Overall, it may be possible to construct a scheme with
five rather than six different parameters. This can be beneficial to reduce the complexity
of the order conditions, and also to reuse storage. 
We can trade $\beta_1$ and $\beta_2$ for
a single coefficient $c$ by requiring:
\begin{eqnarray}
\beta_1=c\alpha_{31},\,\,\,\,\,\beta_2=c\alpha_{32}.\label{eq_betacalpha}
\end{eqnarray}

This way one can store the combination $\alpha_{32}K_2+\alpha_{31}K_1$,
rather than $K_1$ and $K_2$ separately, overwriting $K_1$ with this combination.

As is done for classical RK methods, by Taylor expanding the numerical scheme and comparing
with the expansion of the exact solution one arrives at the following order conditions:
\begin{eqnarray}
\phantom{+\frac{1}{2}\beta_3(\beta_3+c(\alpha_{32}+\alpha_{31}))
(\alpha_{32}+\alpha_{31}+\alpha_{21})}
\mathllap{\beta_3+(1+c)(\alpha_{32}+\alpha_{31})+\alpha_{21}}&=&1,\label{eq_co1}
\end{eqnarray}
\begin{eqnarray}
\phantom{+\frac{1}{2}\beta_3(\beta_3+c(\alpha_{32}+\alpha_{31}))
	(\alpha_{32}+\alpha_{31}+\alpha_{21})}
\mathllap{\beta_3(\alpha_{32}+\alpha_{31}+\alpha_{21})+(1+c)\alpha_{32}\alpha_{21}}
&=&\frac{1}{2},\label{eq_co2}
\end{eqnarray}
\begin{eqnarray}
\phantom{+\frac{1}{2}\beta_3(\beta_3+c(\alpha_{32}+\alpha_{31}))
	(\alpha_{32}+\alpha_{31}+\alpha_{21})}
\mathllap{\beta_3(\alpha_{32}+\alpha_{31}+\alpha_{21})^2
+(1+c)\alpha_{32}\alpha_{21}^2}&=&\frac{1}{3},\label{eq_co3}
\end{eqnarray}
\begin{eqnarray}
\phantom{+\frac{1}{2}\beta_3(\beta_3+c(\alpha_{32}+\alpha_{31}))
	(\alpha_{32}+\alpha_{31}+\alpha_{21})}
\mathllap{\beta_3\alpha_{32}\alpha_{21}}&=&\frac{1}{6},\label{eq_co4}
\end{eqnarray}
\begin{eqnarray}
\frac{3}{2}\left[\beta_3(\alpha_{32}+\alpha_{31}+\alpha_{21})^2
+(1+c)\alpha_{32}\alpha_{21}^2\right]&\phantom{=}&\nonumber\\
&\phantom{=}&\nonumber\\
+\frac{1}{2}\beta_3(\beta_3+c(\alpha_{32}+\alpha_{31}))
(\alpha_{32}+\alpha_{31}+\alpha_{21})
&\phantom{=}&\nonumber\\
+\frac{1}{2}[c(\beta_3+c(\alpha_{32}+\alpha_{31}))\nonumber\\
+\alpha_{32}+\alpha_{31}]
\alpha_{32}\alpha_{21}&\phantom{=}&\nonumber\\
+c(\alpha_{32}+\alpha_{31})
\alpha_{32}\alpha_{21}
&=&\frac{1}{2},
\label{eq_nco1}
\end{eqnarray}
\begin{eqnarray}
\phantom{+\frac{1}{2}\beta_3(\beta_3+c(\alpha_{32}+\alpha_{31}))
	(\alpha_{32}+\alpha_{31}+\alpha_{21})}
\mathllap{\frac{1}{2}\beta_3(\beta_3+c(\alpha_{32}+\alpha_{31}))
(\alpha_{32}+\alpha_{31}+\alpha_{21})}
&\phantom{=}&\nonumber\\
+\frac{1}{2}[c(\beta_3+c(\alpha_{32}+\alpha_{31}))\nonumber\\
+\alpha_{32}+\alpha_{31}]
\alpha_{32}\alpha_{21}
&\phantom{=}&\nonumber\\
+(\beta_3+c(\alpha_{32}+\alpha_{31}))
\alpha_{32}\alpha_{21}
&=&\frac{1}{6}.
\label{eq_nco2}
\end{eqnarray}
Eqs.~(\ref{eq_co1})--(\ref{eq_co4}) are equivalent to the classical order conditions
for a third-order scheme, expressed in terms of the coefficients
$\alpha_{ij}$, $\beta_i$
(their relation to the classical coefficients will become clear in a moment).
Eqs.~(\ref{eq_nco1}), (\ref{eq_nco2}) appear when $Y$ is a vector or a matrix and
terms $F'F^2$ and $FF'F$ in the Taylor expansion can no longer be combined.
By using the four classical order conditions (\ref{eq_co1})--(\ref{eq_co4}) one finds
that the two conditions (\ref{eq_nco1}), (\ref{eq_nco2}) are not independent and can be
condensed into a single condition:
\begin{equation}
\frac{1}{2}\left[\beta_3+c(\alpha_{32}+\alpha_{31})\right]+
(1+c)(\alpha_{32}+\alpha_{31})\alpha_{32}\alpha_{21}=\frac{1}{6}.\label{eq_nco}
\end{equation}
Now there are five unknowns and five (non-linear) equations. As is obvious from 
Eqs.~(\ref{eq_co1})--(\ref{eq_co4}), (\ref{eq_nco}) this system can become significantly
simpler if $1+c=0$, so we can try $c=-1$ as a first guess. Then dividing
(\ref{eq_co3}) by (\ref{eq_co2}) we get 
$\alpha_{32}+\alpha_{31}+\alpha_{21}=2/3$,
substituting that into (\ref{eq_co2}) we immediately get
$\beta_3=3/4$, from (\ref{eq_co1})
$\alpha_{21}=1/4$, 
from (\ref{eq_co4}) $\alpha_{32}=8/9$, 
and, finally, $\alpha_{31}=-17/36$.
We can check that Eq.~(\ref{eq_nco}) is also satisfied. These are nothing else
but the coefficients of the scheme of Ref.~\cite{Luscher:2010iy},
Algorithm~\ref{alg_W6}.

However, as the reader can verify, there are solutions for other values of $c$.
This is possible because, as one can show, in the form involving the
coefficient $c$, Eq.~(\ref{eq_betacalpha}),
the non-classical constraint (\ref{eq_nco}) is a linear combination
of (\ref{eq_co1})--(\ref{eq_co3}). 
Had we kept $\beta_1$ and $\beta_2$ as independent
coefficients this would not happen. We would still end up with a one-parameter family
of solutions (six coefficients with five constraints), but (\ref{eq_nco}) would be
linearly independent.

As is now clear from the discussion in Sec.~\ref{sec_ls_lie}, the integrator
of \cite{Luscher:2010iy} is, in fact, a low-storage commutator-free Lie group
integrator that belongs to the family of schemes based on the
classical $2N$-storage methods of Ref.~\cite{WILLIAMSON198048}.
See~\cite{Bazavov2020} for detailed discussion.
We call this scheme LSCFRK3W6 in Sec.~\ref{sec_order3}.
Its set of classical RK coefficients is
$a_{21}=1/4$, $a_{31}=-2/9$, $a_{32}=8/9$, $b_1=1/4$, $b_2=0$, $b_3=3/4$
and they are related to the set of $\alpha$'s and $\beta$'s we started with as
\begin{eqnarray}
\alpha_{21} &=& a_{21},\\
\alpha_{31} &=& a_{31}-a_{21},\\
\alpha_{32} &=& a_{32},\\
\beta_{1} &=& b_1-a_{31},\\
\beta_{2} &=& b_2-a_{32},\\
\beta_{3} &=& b_3.
\end{eqnarray}
The coefficients in the $2N$-storage format, Algorithm~\ref{alg_CF_2N_RK},
are given in the second column of Table~\ref{tab_2N}.

\section{Mathematica script}
\label{sec_app_math}
A Wolfram Mathematica script (tested with version 11) that calculates
the coefficients for the $2N$-storage explicit three-stage third-order
Runge-Kutta methods from provided $c_2$ and $c_3$ coefficients
is given below.
The reader can simply copy and paste it into an empty Mathematica
notebook (the formatting will most likely be lost).

\begin{verbatim}
(* Note: enclosing parentheses are needed
so that Abort[] function could actually abort
the execution of the cell *)
( (* <---- do not remove *)
(* SET c2 AND c3 HERE e.g. from Table B.2 *)
c2 := 1/4;
c3 := 2/3;

(* check the singular point *)
If[c2 == 1/3 && c3 == 1/3, 
Print["No RK scheme with c2=c3=1/3 exists"];
Abort[]];

(* check if low-storage *)
If[c3^2*(1 - c2) + c3*(c2^2 + 1/2*c2 - 1)
+ (1/3 - 1/2*c2) != 0, 
Print["c2,c3 -- not a low-storage scheme"];
Abort[]];

(* check if limiting case c2=2/3,c3=0
or c2=c3=2/3 is hit *)

If[c2 == 2/3 && c3 == 0, 
b3 := -1/3; b2 := 3/4; b1 := 1/4 - b3; 
a32 := 1/4/b3; a31 := -a32; a21 := 2/3, 
If[c2 == 2/3 && c3 == 2/3,
b3 := 1/3; b2 := 3/4 - b3; b1 := 1/4; 
a32 := 1/4/b3; a31 := 2/3 - a32; a21 := 2/3, 
b2 := (3*c3 - 2)/6/c2/(c3 - c2);
b3 := (2 - 3*c2)/6/c3/(c3 - c2); 
a32 := c3*(c3 - c2)/c2/(2 - 3*c2);
b1 := 1 - b2 - b3; a31 = c3 - a32; a21 := c2]];

Print["Coefficients in classical RK form:"];
Print["a21=", a21];
Print["a31=", a31];
Print["a32=", a32];
Print["b1=", b1];
Print["b2=", b2];
Print["b3=", b3];

alpha21 := a21;
alpha31 := a31 - a21;
alpha32 := a32;
beta3 := b3;
c := (b1 - a31)/(a31 - a21);

Print["Coefficients in the form of"];
Print["Luescher, 1006.4518"];
Print["with reusability condition"];
Print["beta1=c*alpha31, beta2=c*alpha32:"];
Print["alpha21=", alpha21];
Print["alpha31=", alpha31];
Print["alpha32=", alpha32];
Print["beta3=", beta3];
Print["c=", c];

A1 := 0;
B3 := b3;
B2 := a32;
A3 := (b2 - B2)/b3;
B1 := a21;
If[b2 == 0,
A2 := (a31 - a21)/a32, A2 := (b1 - B1)/b2];

Print["Low-storage form of Williamson:"];
Print["A1=", A1];
Print["A2=", A2];
Print["A3=", A3];
Print["B1=", B1];
Print["B2=", B2];
Print["B3=", B3];

Print["Variable step size:"];
Print["Second-order coefficients"];
Ds := c2*alpha32 - c3*(c3 - c2);
If[Ds == 0,
Print["No reusable embedded scheme"];
Print["Pick lambda3=0"];
l3 := 0;
l2 := 1/2/c2;
l1 := 1 - l2;
Print["lambda1=", l1];
Print["lambda2=", l2];
Print["lambda3=", l3],
l2 := alpha32*(1/2 - c3)/Ds;
l3 := 1 - (c3 - c2)*(1/2 - c3)/Ds;
l1 := 1 - l2 - l3;
q := l2/a32;
Print["with reuse of third-order second stage"];
Print["lambda2=q*alpha32, lambda1=q*alpha31"];
Print["lambda3=", l3];
Print["q=", q]];
) (* <---- do not remove *)
(* end of script *)
\end{verbatim}





\bibliography{eff_gflow}

\begin{thebibliography}{10}
\expandafter\ifx\csname url\endcsname\relax
  \def\url#1{\texttt{#1}}\fi
\expandafter\ifx\csname urlprefix\endcsname\relax\def\urlprefix{URL }\fi
\expandafter\ifx\csname href\endcsname\relax
  \def\href#1#2{#2} \def\path#1{#1}\fi

\bibitem{Wilson:1974sk}
K.~G. Wilson, {Confinement of Quarks}, Phys. Rev. D10 (1974) 2445--2459.
\newblock \href {https://doi.org/10.1103/PhysRevD.10.2445}
  {\path{doi:10.1103/PhysRevD.10.2445}}.

\bibitem{Luscher:2010iy}
M.~Luescher, {Properties and uses of the {Wilson} flow in lattice QCD}, JHEP 08
  (2010) 071, [Erratum: JHEP03,092(2014)].
\newblock \href {http://arxiv.org/abs/1006.4518} {\path{arXiv:1006.4518}},
  \href {https://doi.org/10.1007/JHEP08(2010)071, 10.1007/JHEP03(2014)092}
  {\path{doi:10.1007/JHEP08(2010)071, 10.1007/JHEP03(2014)092}}.

\bibitem{Hairer2006}
E.~Hairer, C.~Lubich, G.~Wanner,
  \href{https://cds.cern.ch/record/1250576}{{Geometric Numerical Integration:
  Structure-Preserving Algorithms for Ordinary Differential Equations; 2nd
  ed.}}, Springer, Dordrecht, 2006.
\newblock \href {https://doi.org/10.1007/3-540-30666-8}
  {\path{doi:10.1007/3-540-30666-8}}.
\newline\urlprefix\url{https://cds.cern.ch/record/1250576}

\bibitem{Bazavov2020}
A.~{Bazavov}, {Commutator-free Lie group methods with minimum storage
  requirements and reuse of exponentials}, arXiv e-prints (2020)
  arXiv:2007.04225\href {http://arxiv.org/abs/2007.04225}
  {\path{arXiv:2007.04225}}.

\bibitem{Luscher:1985zq}
M.~Luscher, P.~Weisz, {Computation of the Action for On-Shell Improved Lattice
  Gauge Theories at Weak Coupling}, Phys. Lett. 158B (1985) 250--254.
\newblock \href {https://doi.org/10.1016/0370-2693(85)90966-9}
  {\path{doi:10.1016/0370-2693(85)90966-9}}.

\bibitem{Borsanyi:2012zs}
S.~Borsanyi, et~al., {High-precision scale setting in lattice QCD}, JHEP 09
  (2012) 010.
\newblock \href {http://arxiv.org/abs/1203.4469} {\path{arXiv:1203.4469}},
  \href {https://doi.org/10.1007/JHEP09(2012)010}
  {\path{doi:10.1007/JHEP09(2012)010}}.

\bibitem{Fodor:2014cpa}
Z.~Fodor, K.~Holland, J.~Kuti, S.~Mondal, D.~Nogradi, C.~H. Wong, {The lattice
  gradient flow at tree-level and its improvement}, JHEP 09 (2014) 018.
\newblock \href {http://arxiv.org/abs/1406.0827} {\path{arXiv:1406.0827}},
  \href {https://doi.org/10.1007/JHEP09(2014)018}
  {\path{doi:10.1007/JHEP09(2014)018}}.

\bibitem{ButcherBook}
J.~Butcher, Numerical Methods for Ordinary Differential Equations, 3rd Edition,
  Wiley, 2016.

\bibitem{HairerBook1}
E.~Hairer, S.~N{\o}rsett, G.~Wanner, Solving Ordinary Differential Equations
  {I} Nonstiff problems, 2nd Edition, Springer, Berlin, 2000.

\bibitem{WILLIAMSON198048}
J.~Williamson,
  \href{http://www.sciencedirect.com/science/article/pii/0021999180900339}{Low-storage
  {Runge}-{Kutta} schemes}, Journal of Computational Physics 35~(1) (1980) 48
  -- 56.
\newblock \href {https://doi.org/https://doi.org/10.1016/0021-9991(80)90033-9}
  {\path{doi:https://doi.org/10.1016/0021-9991(80)90033-9}}.
\newline\urlprefix\url{http://www.sciencedirect.com/science/article/pii/0021999180900339}

\bibitem{KENNEDY2000177}
C.~A. Kennedy, M.~H. Carpenter, R.~Lewis,
  \href{http://www.sciencedirect.com/science/article/pii/S0168927499001415}{Low-storage,
  explicit {Runge}-{Kutta} schemes for the compressible {Navier}-{Stokes}
  equations}, Applied Numerical Mathematics 35~(3) (2000) 177 -- 219.
\newblock \href {https://doi.org/https://doi.org/10.1016/S0168-9274(99)00141-5}
  {\path{doi:https://doi.org/10.1016/S0168-9274(99)00141-5}}.
\newline\urlprefix\url{http://www.sciencedirect.com/science/article/pii/S0168927499001415}

\bibitem{KETCHESON20101763}
D.~I. Ketcheson,
  \href{http://www.sciencedirect.com/science/article/pii/S0021999109006251}{{Runge}-{Kutta}
  methods with minimum storage implementations}, Journal of Computational
  Physics 229~(5) (2010) 1763 -- 1773.
\newblock \href {https://doi.org/https://doi.org/10.1016/j.jcp.2009.11.006}
  {\path{doi:https://doi.org/10.1016/j.jcp.2009.11.006}}.
\newline\urlprefix\url{http://www.sciencedirect.com/science/article/pii/S0021999109006251}

\bibitem{CK1994}
M.~Carpenter, C.~Kennedy, Fourth-order {2N}-storage {Runge}-{Kutta} schemes,
  Tech. Rep. NASA-TM-109112, NASA (1994).

\bibitem{BERNARDINI20094182}
M.~Bernardini, S.~Pirozzoli,
  \href{http://www.sciencedirect.com/science/article/pii/S0021999109001077}{A
  general strategy for the optimization of {Runge}-{Kutta} schemes for wave
  propagation phenomena}, Journal of Computational Physics 228~(11) (2009) 4182
  -- 4199.
\newblock \href {https://doi.org/https://doi.org/10.1016/j.jcp.2009.02.032}
  {\path{doi:https://doi.org/10.1016/j.jcp.2009.02.032}}.
\newline\urlprefix\url{http://www.sciencedirect.com/science/article/pii/S0021999109001077}

\bibitem{BERLAND20061459}
J.~Berland, C.~Bogey, C.~Bailly,
  \href{http://www.sciencedirect.com/science/article/pii/S0045793005000575}{Low-dissipation
  and low-dispersion fourth-order {Runge}-{Kutta} algorithm}, Computers and
  Fluids 35~(10) (2006) 1459 -- 1463.
\newblock \href
  {https://doi.org/https://doi.org/10.1016/j.compfluid.2005.04.003}
  {\path{doi:https://doi.org/10.1016/j.compfluid.2005.04.003}}.
\newline\urlprefix\url{http://www.sciencedirect.com/science/article/pii/S0045793005000575}

\bibitem{TOULORGE20122067}
T.~Toulorge, W.~Desmet,
  \href{http://www.sciencedirect.com/science/article/pii/S0021999111006796}{Optimal
  {Runge}-{Kutta} schemes for discontinuous {Galerkin} space discretizations
  applied to wave propagation problems}, Journal of Computational Physics
  231~(4) (2012) 2067 -- 2091.
\newblock \href {https://doi.org/https://doi.org/10.1016/j.jcp.2011.11.024}
  {\path{doi:https://doi.org/10.1016/j.jcp.2011.11.024}}.
\newline\urlprefix\url{http://www.sciencedirect.com/science/article/pii/S0021999111006796}

\bibitem{STANESCU1998674}
D.~Stanescu, W.~Habashi,
  \href{http://www.sciencedirect.com/science/article/pii/S0021999198959861}{{2N}-storage
  low dissipation and dispersion {Runge}-{Kutta} schemes for computational
  acoustics}, Journal of Computational Physics 143~(2) (1998) 674 -- 681.
\newblock \href {https://doi.org/https://doi.org/10.1006/jcph.1998.5986}
  {\path{doi:https://doi.org/10.1006/jcph.1998.5986}}.
\newline\urlprefix\url{http://www.sciencedirect.com/science/article/pii/S0021999198959861}

\bibitem{ALLAMPALLI20093837}
V.~Allampalli, R.~Hixon, M.~Nallasamy, S.~D. Sawyer,
  \href{http://www.sciencedirect.com/science/article/pii/S0021999109000849}{High-accuracy
  large-step explicit {Runge}-{Kutta} ({HALE-RK}) schemes for computational
  aeroacoustics}, Journal of Computational Physics 228~(10) (2009) 3837 --
  3850.
\newblock \href {https://doi.org/https://doi.org/10.1016/j.jcp.2009.02.015}
  {\path{doi:https://doi.org/10.1016/j.jcp.2009.02.015}}.
\newline\urlprefix\url{http://www.sciencedirect.com/science/article/pii/S0021999109000849}

\bibitem{NIEGEMANN2012364}
J.~Niegemann, R.~Diehl, K.~Busch,
  \href{http://www.sciencedirect.com/science/article/pii/S0021999111005213}{Efficient
  low-storage {Runge}-{Kutta} schemes with optimized stability regions},
  Journal of Computational Physics 231~(2) (2012) 364 -- 372.
\newblock \href {https://doi.org/https://doi.org/10.1016/j.jcp.2011.09.003}
  {\path{doi:https://doi.org/10.1016/j.jcp.2011.09.003}}.
\newline\urlprefix\url{http://www.sciencedirect.com/science/article/pii/S0021999111005213}

\bibitem{Yan2017}
Y.~an~Yan, Low-storage {Runge}-{Kutta} method for simulating time-dependent
  quantum dynamics, Chinese Journal of Chemical Physics 30~(3) (2017) 277 --
  286.

\bibitem{MUNTHEKAAS1999115}
H.~Munthe-Kaas,
  \href{http://www.sciencedirect.com/science/article/pii/S0168927498000300}{High
  order {Runge}-{Kutta} methods on manifolds}, Applied Numerical Mathematics
  29~(1) (1999) 115 -- 127, proceedings of the NSF/CBMS Regional Conference on
  Numerical Analysis of Hamiltonian Differential Equations.
\newblock \href {https://doi.org/https://doi.org/10.1016/S0168-9274(98)00030-0}
  {\path{doi:https://doi.org/10.1016/S0168-9274(98)00030-0}}.
\newline\urlprefix\url{http://www.sciencedirect.com/science/article/pii/S0168927498000300}

\bibitem{MuntheKaasOwren1999}
H.~Z. Munthe-Kaas, B.~Owren, Computations in a free {Lie} algebra,
  Philosophical Transactions of the Royal Society of London. Series A:
  Mathematical, Physical and Engineering Sciences 357 (1999) 957 -- 981.

\bibitem{Crouch1993}
P.~E. Crouch, R.~Grossman, \href{https://doi.org/10.1007/BF02429858}{Numerical
  integration of ordinary differential equations on manifolds}, Journal of
  Nonlinear Science 3~(1) (1993) 1--33.
\newblock \href {https://doi.org/10.1007/BF02429858}
  {\path{doi:10.1007/BF02429858}}.
\newline\urlprefix\url{https://doi.org/10.1007/BF02429858}

\bibitem{CELLEDONI2003341}
E.~Celledoni, A.~Marthinsen, B.~Owren,
  \href{http://www.sciencedirect.com/science/article/pii/S0167739X02001619}{Commutator-free
  {Lie} group methods}, Future Generation Computer Systems 19~(3) (2003) 341 --
  352, special Issue on Geometric Numerical Algorithms.
\newblock \href {https://doi.org/https://doi.org/10.1016/S0167-739X(02)00161-9}
  {\path{doi:https://doi.org/10.1016/S0167-739X(02)00161-9}}.
\newline\urlprefix\url{http://www.sciencedirect.com/science/article/pii/S0167739X02001619}

\bibitem{Knoth2020git}
O.~Knoth, \href{https://github.com/OsKnoth/B-Series}{{B-Series}} (2020).
\newline\urlprefix\url{https://github.com/OsKnoth/B-Series}

\bibitem{Follana:2006rc}
E.~Follana, Q.~Mason, C.~Davies, K.~Hornbostel, G.~P. Lepage, J.~Shigemitsu,
  H.~Trottier, K.~Wong, {Highly improved staggered quarks on the lattice, with
  applications to charm physics}, Phys. Rev. D75 (2007) 054502.
\newblock \href {http://arxiv.org/abs/hep-lat/0610092}
  {\path{arXiv:hep-lat/0610092}}, \href
  {https://doi.org/10.1103/PhysRevD.75.054502}
  {\path{doi:10.1103/PhysRevD.75.054502}}.

\bibitem{Bazavov:2010ru}
A.~Bazavov, et~al., {Scaling studies of QCD with the dynamical HISQ action},
  Phys. Rev. D82 (2010) 074501.
\newblock \href {http://arxiv.org/abs/1004.0342} {\path{arXiv:1004.0342}},
  \href {https://doi.org/10.1103/PhysRevD.82.074501}
  {\path{doi:10.1103/PhysRevD.82.074501}}.

\bibitem{Bazavov:2012xda}
A.~Bazavov, et~al., {Lattice QCD Ensembles with Four Flavors of Highly Improved
  Staggered Quarks}, Phys. Rev. D87~(5) (2013) 054505.
\newblock \href {http://arxiv.org/abs/1212.4768} {\path{arXiv:1212.4768}},
  \href {https://doi.org/10.1103/PhysRevD.87.054505}
  {\path{doi:10.1103/PhysRevD.87.054505}}.

\bibitem{Bazavov:2015yea}
A.~Bazavov, et~al., {Gradient flow and scale setting on {MILC} {HISQ}
  ensembles}, Phys. Rev. D 93~(9) (2016) 094510.
\newblock \href {http://arxiv.org/abs/1503.02769} {\path{arXiv:1503.02769}},
  \href {https://doi.org/10.1103/PhysRevD.93.094510}
  {\path{doi:10.1103/PhysRevD.93.094510}}.

\bibitem{PRINCE198167}
P.~Prince, J.~Dormand,
  \href{http://www.sciencedirect.com/science/article/pii/0771050X81900103}{High
  order embedded runge-kutta formulae}, Journal of Computational and Applied
  Mathematics 7~(1) (1981) 67 -- 75.
\newblock \href {https://doi.org/https://doi.org/10.1016/0771-050X(81)90010-3}
  {\path{doi:https://doi.org/10.1016/0771-050X(81)90010-3}}.
\newline\urlprefix\url{http://www.sciencedirect.com/science/article/pii/0771050X81900103}

\bibitem{Ralston1962}
A.~Ralston,
  \href{https://www.ams.org/journals/mcom/1962-16-080/S0025-5718-1962-0150954-0/}{{Runge}-{Kutta}
  methods with minimum error bounds}, Math. Comp. 16 (1962) 431 -- 437.
\newblock \href
  {https://doi.org/https://doi.org/10.1090/S0025-5718-1962-0150954-0}
  {\path{doi:https://doi.org/10.1090/S0025-5718-1962-0150954-0}}.
\newline\urlprefix\url{https://www.ams.org/journals/mcom/1962-16-080/S0025-5718-1962-0150954-0/}

\bibitem{Fritzsch:2013je}
P.~Fritzsch, A.~Ramos, {The gradient flow coupling in the Schroedinger
  Functional}, JHEP 10 (2013) 008.
\newblock \href {http://arxiv.org/abs/1301.4388} {\path{arXiv:1301.4388}},
  \href {https://doi.org/10.1007/JHEP10(2013)008}
  {\path{doi:10.1007/JHEP10(2013)008}}.

\bibitem{WandeltGuenther2016}
M.~Wandelt, M.~Guenther, Efficient numerical simulation of the wilson flow in
  lattice qcd, in: G.~Russo, V.~Capasso, G.~Nicosia, V.~Romano (Eds.), Progress
  in Industrial Mathematics at ECMI 2014, Springer International Publishing,
  Cham, 2016, pp. 1065--1071.

\bibitem{BOGACKI1989321}
P.~Bogacki, L.~Shampine,
  \href{http://www.sciencedirect.com/science/article/pii/0893965989900797}{A
  3(2) pair of runge - kutta formulas}, Applied Mathematics Letters 2~(4)
  (1989) 321 -- 325.
\newblock \href {https://doi.org/https://doi.org/10.1016/0893-9659(89)90079-7}
  {\path{doi:https://doi.org/10.1016/0893-9659(89)90079-7}}.
\newline\urlprefix\url{http://www.sciencedirect.com/science/article/pii/0893965989900797}

\bibitem{Ce:2015qha}
M.~Ce, C.~Consonni, G.~P. Engel, L.~Giusti, {Non-Gaussianities in the
  topological charge distribution of the SU(3) Yang--Mills theory}, Phys. Rev.
  D92~(7) (2015) 074502.
\newblock \href {http://arxiv.org/abs/1506.06052} {\path{arXiv:1506.06052}},
  \href {https://doi.org/10.1103/PhysRevD.92.074502}
  {\path{doi:10.1103/PhysRevD.92.074502}}.

\bibitem{MILCgit}
{MILC Collaboration}, \href{https://github.com/milc-qcd/milc_qcd}{{MILC
  collaboration code for lattice QCD calculations}} (Dec. 2020).
\newline\urlprefix\url{https://github.com/milc-qcd/milc_qcd}

\bibitem{MuntheKaas1998}
H.~Munthe-Kaas, \href{https://doi.org/10.1007/BF02510919}{{Runge}-{Kutta}
  methods on {Lie} groups}, BIT Numerical Mathematics 38~(1) (1998) 92--111.
\newblock \href {https://doi.org/10.1007/BF02510919}
  {\path{doi:10.1007/BF02510919}}.
\newline\urlprefix\url{https://doi.org/10.1007/BF02510919}

\end{thebibliography}




\end{document}